\documentclass[3p]{elsarticle}

\usepackage{amsmath,amssymb}
\usepackage{graphicx}
\usepackage{bm}
\usepackage{booktabs}
\usepackage[hidelinks]{hyperref}

\biboptions{sort&compress}
\journal{Annals of Physics}

\newlength{\onecolfigwidth}
\AtBeginDocument{\setlength{\onecolfigwidth}{\ifdim\columnwidth<\textwidth\columnwidth\else0.6\textwidth\fi}}

\begin{document}

\begin{frontmatter}

\title{The electromagnetic energy--momentum tensor in dielectrics: What force and work decide}

\author{Bela Schulte Westhoff}
\ead{schultewesthoff@tu-berlin.de}
\affiliation{organization={Technische Universit{\"a}t Berlin},
            postcode={10587},
            city={Berlin},
            country={Germany}}

\begin{abstract}
The electromagnetic energy--momentum tensor in matter has been debated for over a century. The prevailing resolutions treat the division between field and matter as a convention, or accept both the Minkowski and Abraham momenta. For the electric response of matter in classical electrodynamics, we argue that the division is not a convention. Such a tensor asserts a force on matter and a power. We require, as mechanics does, that the power be the work of that force at the velocity of the mass that receives it, and that this hold for every material and motion. Of the tensors written in the macroscopic fields, one meets both requirements, and it is known: the vacuum-form tensor of $\mathbf E$ and $\mathbf B$. These fields are part of the microscopic field, and the tensor gives their Lorentz force on all charge, bound and free, and its work at the velocities of the charges. Total force and torque follow from the fields alone; heat and the load that deforms a body follow together with laws of the material: for a linear dielectric the Helmholtz stress, electrostriction included. Minkowski's and Abraham's tensors count the polarization as a field of its own. Even for linear response they omit electrostriction, and when the body deforms their power contains work that their force does not do; for nonlinear or lossy response at rest they have a power and no work. Equal totals and the kinetic--canonical correspondence remain valid but select no tensor; the debate's experiments agree with the Lorentz force.
\end{abstract}


\end{frontmatter}

\section{Introduction}
\label{sec:introduction}

Electromagnetic fields in matter are engineered with great precision: every electronic and optical device depends on them, and the macroscopic Maxwell equations predict them reliably. Yet the most elementary questions about these fields have no agreed answer. How much energy and momentum does the field hold inside matter? How do they flow? What force does the field exert on the matter, and what work does it do? Together these questions ask for the energy--momentum tensor of the field. In vacuum it is not in doubt. In matter its form has been debated for more than a century~\cite{Pfeifer2007,MilonniBoyd2010,Kemp2015MacroscopicTheoryOpticalMomentum,AnghinoniEtAl2022}.

The tensor has four entries: the energy density $u$ and the energy flux $\mathbf S$, the momentum density $\mathbf g$ and the stress $\mathbf T$. Its two balances give what the field hands to matter, a force density and a power density:
\begin{equation}
    \mathbf f=\nabla\cdot\mathbf T-\partial_t\mathbf g,
    \qquad
    P=-\left(\partial_tu+\nabla\cdot\mathbf S\right).
\label{eq:intro-two-balances}
\end{equation}
In mechanics the two are not independent: the power of a force is its work, done at the velocity of the mass on which it acts. The debate is best known for one entry, the momentum of light in a medium~\cite{Pfeifer2007,MilonniBoyd2010,Barnett2010}: in their conventional forms~\cite[Eqs.~(40) and (42)]{Pfeifer2007}, the Minkowski~\cite{Minkowski1908} and Abraham~\cite{Abraham1909,Abraham1910} tensors share $u$ and $\mathbf S$ and differ in $\mathbf g$ and $\mathbf T$. But the whole tensor is in question. Further tensors have been proposed besides these two, among them the vacuum-form tensor of the Amperian formulation~\cite{Kemp2015MacroscopicTheoryOpticalMomentum,AnghinoniEtAl2022}; they differ in the energy density as well, and with it in the power. The search for an answer has nevertheless concentrated on the first balance, the force~\cite[Sec.~1]{Saldanha2011}. What is measured is a force or a torque on a body, or the displacement or deformation that results~\cite[Sec.~III]{Pfeifer2007}, and it is compared with the force density of each candidate~\cite{Kemp2015MacroscopicTheoryOpticalMomentum,AnghinoniEtAl2022}.

Measurements have not settled it. For steady fields the total force and torque on a body are the same for every candidate tensor~\cite{BarnettLoudon2006ForceDielectricMedium}. In time-dependent fields they differ~\cite{Markel2020Balazs}, and a measured torque has been read as excluding one of them~\cite{WalkerLahozWalker1975}; each candidate, however, comes with a material counterpart that restores the total~\cite{BrevikEllingsen2011,BrevikEllingsen2012Reply}. Inside the body the force densities differ, and measured deformations are expected to decide between them~\cite{Brevik1979,MansuripurZakharianWright2013}. So far, optical experiments have been read in favor of either momentum, depending on the mechanical response of the medium~\cite{Leonhardt2014,ZhangShePengLeonhardt2015}, or found to have no bearing on the tensor~\cite{Brevik2017,VermaBrevikLi2025}.

The prevailing resolution holds that measurements cannot settle it. An electromagnetic tensor is never tested alone: each one, taken with its own material counterpart, gives the same total and the same predictions, so that the division of the total into an electromagnetic and a material part is a convention~\cite[Secs.~7.6 and 7.7]{PenfieldHaus1967}, \cite[Chap.~V, Sec.~7]{DeGrootSuttorp1972}; see also Refs.~\cite{Kranys1979EquivalenceAbrahamMinkowski,Kranys1982NonUniqueness,Maugin1980EquivalenceAbrahamMinkowskiOthers,Pfeifer2007,RamosRubilarObukhov2015FirstPrinciples,Silveirinha2017ReexaminationAbrahamMinkowski}. The kinetic--canonical resolution rests on the same freedom of division: it identifies the Abraham and Minkowski momenta with the kinetic and the canonical momentum of light, so that both are correct in their own roles~\cite{Barnett2010,BarnettLoudon2010EnigmaOpticalMomentum}. Single tensors have also been argued for, the vacuum-form tensor of the macroscopic $\mathbf E$ and $\mathbf B$ with the Lorentz force on all charge and current~\cite{Nelson1991,ObukhovHehl2003EnergyMomentumForcesMatter,Saldanha2011,AnghinoniPartanenAstrath2024LocalModel} and others~\cite{Kemp2015MacroscopicTheoryOpticalMomentum,Mansuripur2010,MedinaStephany2014,AnghinoniPartanenAstrath2023}, but none has become the accepted answer. After more than a century, the prevailing answer to the questions above is thus that the tensor is a convention, or that two momenta are both correct.

We argue the opposite: the division is not a convention. The debate has asked how the total is to be divided between field and matter, and the totals leave this open. They leave it open for a free charge as well, and no one calls the force on a free charge a matter of choice. We ask instead what mechanics asks of any force: which mass receives it, at which velocity it does its work, what balances it, and where the work goes. We start from the textbook definition of the macroscopic fields, the average of the microscopic field~\cite[Sec.~6.6]{Jackson1999}: a compressed description of it, and a part of it. We read force and power together, where the debate has compared force densities: the load that deforms a body does not follow from the force alone. We require mechanical consistency of the tensor as a whole: its power must be the work of its own force, at the velocity of the mass that receives it. And we require generality: this must hold for the exchange with both continuum mechanics and thermodynamics, and for every material, linear or nonlinear, reversible or lossy, at rest or in motion. The answers are read from the exact identities of the microscopic theory, with the following results.
\begin{enumerate}
\item \emph{One tensor is that of the field.} The macroscopic fields $\mathbf E$ and $\mathbf B$ exert their share of the Lorentz force on the ions and the electrons, whatever binds them. Their energy--momentum tensor is therefore the vacuum-form tensor: its two balances, Eq.~\eqref{eq:intro-two-balances}, are $\mathbf f=\rho\mathbf E+\mathbf J\times\mathbf B$ and $P=\mathbf J\cdot\mathbf E$, with the total charge and current, bound and free. The second is the work of the first, at the velocities of the charges. No law of the material enters, so the tensor is the same for every material and every motion, and it keeps its symmetric, covariant form in every inertial frame (Secs.~\ref{sec:microscopic-baseline}, \ref{sec:macroscopic-compression}, and \ref{subsec:moving-source-geometry}).
\item \emph{Force and power together lead to the load on a body.} The total force and torque on a body follow from the fields alone. The load that deforms it follows from the two balances together, with a law of the material~\cite{SchulteWesthoffLocalLoads2026,SchulteWesthoffDielectricEnergy2026}: the force is the share that the fields exert on the charges directly, the power records the energy passed to the matter, and the change of its stored part under deformation completes the load, as in the Helmholtz stress of a linear dielectric, electrostriction included (Sec.~\ref{subsec:mechanical-load-closure}). A comparison of force densities written in the fields alone therefore cannot decide between the tensors.
\item \emph{Every part of the energy and momentum has an owner.} Once $\mathbf E$ and $\mathbf B$ are defined, the total energy and momentum of field and matter divide exactly into three parts: those of the macroscopic fields, of the field structure that averaging removes, and of the ions and the electrons. The force and the work of the fields are the budget on which continuum mechanics and thermodynamics draw: the work goes into stored energy, into the motion and deformation of the body, and into heat, in amounts that the laws of the material fix. The three descriptions thus form one map, in which every exchange is a force on an identified mass and its work: a common mechanical foundation for electromagnetic multiphysics (Sec.~\ref{sec:mechanical-unification}).
\item \emph{The other tensors contain a law of the material.} The Minkowski and Abraham tensors take the polarization as a field of its own, although the fields and their sources do not fix it, and thereby add a law of the material to the vacuum-form tensor. They fail the requirements for every response of a dielectric. For linear response, their best case, they give one load where the materials have many, without electrostriction, and when the body deforms their power contains the work of a force that they do not exert. For nonlinear or lossy response they have a power while their force does no work, even in a body at rest. Between resolved charges both put force and power into empty space. What any other candidate adds belongs likewise to the matter, or to no part of the total. The equivalence of the totals and the kinetic--canonical correspondence remain valid and select no tensor, and the experiments of the debate measured forces that the Lorentz force explains (Sec.~\ref{sec:tensor-splits} and \ref{app:separated-charge-support}).
\end{enumerate}

The contrast with the prevailing view is sharp: the division between field and matter is not chosen but found. The energy and the momentum of the field, its force, and its work are definite. They are the basic quantities of macroscopic electrodynamics, every coupling to mechanics and to thermodynamics rests on them, and a coupled model can take them from the fields in the same form for every material.

The result is one of identification, not of prediction. The tensor and its identities are known~\cite{Nelson1991,ObukhovHehl2003EnergyMomentumForcesMatter,Saldanha2011,DatsyukPavlyniuk2015,AnghinoniPartanenAstrath2024LocalModel}, the experiments of the debate agree with it, and no new equation or experiment is needed; what is new is the standard and what it decides. The momentum balance of the tensor has been derived for a general dielectric~\cite{Nelson1991}, and the tensor has been proposed as the natural division, with the Lorentz force on the bound charge~\cite{ObukhovHehl2003EnergyMomentumForcesMatter} and, for linear media without losses, with the energy of the charges counted as that of the matter~\cite{Saldanha2011}. In other treatments that connect the two balances, the force is taken to act on the material as a whole~\cite[Chap.~4]{PenfieldHaus1967} or on atoms treated as dipoles~\cite{DeGrootSuttorp1972,MedinaStephany2014}, and that the field acts on bound charge as on free charge has been adopted as a hypothesis, with opposite conclusions for the momentum of light~\cite{Testa2013,TorchiginTorchigin2014}. To our knowledge, the agreement of force and work on the charges themselves has not been used to decide between the tensors. The argument is classical and concerns the electric response of matter, and ``every material'' is meant within this scope; magnetization and intrinsic spin, where hidden momentum enters the discussion~\cite{Kemp2015MacroscopicTheoryOpticalMomentum,Griffiths2012Comment,KholmetskiiMissevitchYarman2016,CorreaSaldanha2020}, and fluctuation-induced forces are left to separate work.

The paper follows the four questions of mechanics asked above. Section~\ref{sec:microscopic-baseline} answers them where nothing has to be chosen, for the microscopic field and its carriers, and states the two requirements that follow. Section~\ref{sec:tensor-role} turns to the macroscopic fields by averaging and answers the first two questions for every material, with the vacuum-form tensor of $\mathbf E$ and $\mathbf B$. Sections~\ref{sec:dielectric-fecc-examples} and \ref{sec:moving-matter-continuum-work} answer the other two for dielectrics, at rest and in motion, up to the heat and the load. Section~\ref{sec:mechanical-unification} assembles the map of the three descriptions. Section~\ref{sec:tensor-splits} puts the same questions to the Minkowski and Abraham tensors and to the prevailing resolutions, and turns to the experiments. Tables~\ref{tab:mechanical-content}--\ref{tab:tensor-comparison} collect the results: the field identities, the map, and the three tensors side by side.

\section{The mechanical content of the field tensor}
\label{sec:microscopic-baseline}

We start where nothing has to be chosen. Before any averaging there is one field, the microscopic field $\mathbf e$ and $\mathbf b$, and one law of force, and the energy--momentum tensor of that field is not in doubt. Its field expressions carry mechanical information. They say where charged mass is, how it moves, what force it receives, and what work is done on it (Table~\ref{tab:mechanical-content}). This section reads that information out of three exact identities of Maxwell--Lorentz theory. The identities are in every textbook~\cite{Jackson1999,FanoChuAdler1960}; what matters here is what they assert about matter, because a tensor must have this content to be the energy--momentum tensor of a field.

\begin{table}[t]
\caption{What the field expressions say about matter. Each row is an exact identity of Maxwell--Lorentz theory. $\mathbf T$, $\mathbf g$, $u$, and $\mathbf S$ denote the stress, momentum density, energy density, and energy flux of the field.}
\label{tab:mechanical-content}
\renewcommand{\arraystretch}{1.25}
\centering
\begin{tabular}{@{}lp{0.44\columnwidth}l@{}}
\toprule
Field expression
& \raggedright What it says about matter
& Eq. \tabularnewline
\midrule
$\varepsilon_0\nabla\cdot\mathbf e$
& \raggedright where charged mass is
& \eqref{eq:mic-sources-from-fields} \tabularnewline
$\nabla\times\mathbf b/\mu_0-\varepsilon_0\partial_t\mathbf e$
& \raggedright how charged mass moves
& \eqref{eq:mic-sources-from-fields} \tabularnewline
$\nabla\cdot\mathbf T-\partial_t\mathbf g$
& \raggedright the force that charged mass receives
& \eqref{eq:mic-closed-momentum-ledger} \tabularnewline
$-(\partial_tu+\nabla\cdot\mathbf S)$
& \raggedright the work of that force, at the velocity of the mass that receives it
& \eqref{eq:mic-two-sided-energy-ledger} \tabularnewline
\bottomrule
\end{tabular}
\end{table}

An energy--momentum tensor is a statement about mechanics. Its two balances, Eq.~\eqref{eq:intro-two-balances}, assert a force and a power, and in mechanics a force acts on a mass, and its work is that force times the velocity of that mass. Of any tensor of a field one can therefore ask:
\begin{enumerate}
\item[(i)] Which mass receives the force that its momentum balance assigns?
\item[(ii)] At which velocity does that force do the work in its energy balance?
\item[(iii)] What balances the force?
\item[(iv)] Where does the work go?
\end{enumerate}
Questions (i) and (ii) connect the force with the power; (iii) and (iv) say what the matter does with them. For the microscopic field all four have definite answers (Sec.~\ref{subsec:fecc}).

We work in the model of Ref.~\cite[Sec.~2.1]{SchulteWesthoffDielectricEnergy2026}: electrons and ions, which carry charge and mass and which we call \emph{carriers}, with their electromagnetic fields, in the nonrelativistic regime, without intrinsic spin and its magnetic moments. Section~\ref{sec:tensor-role} puts the same questions to the macroscopic fields.

\subsection{Charge sits on mass}
\label{subsec:mic-carrier-ownership}

Each carrier has both charge and mass. Its motion therefore determines an electric current and a mechanical momentum at once.

For each species $a$, the number density $n_a(\mathbf x,t)$ and the velocity $\mathbf v_a(\mathbf x,t)$ describe the distribution and motion of its carriers. With mass $m_a>0$ and charge $q_a\ne0$ per carrier, the electromagnetic sources and the mechanical densities of the species are
\begin{align}
    \rho_a &= q_a n_a,
    & \mathbf j_a &= \rho_a\mathbf v_a,
    \nonumber\\
    \rho_{m,a} &= m_a n_a,
    & \mathbf g_{{\rm mech},a}^{\rm mic}
    &= \rho_{m,a}\mathbf v_a .
    \label{eq:mic-carrier-map}
\end{align}
Summing over species gives
\begin{align}
    \rho_{\rm mic} &= \sum_a\rho_a,
    & \mathbf j_{\rm mic} &= \sum_a\mathbf j_a,
    \nonumber\\
    \mathbf g_{\rm mech}^{\rm mic}
    &= \sum_a\rho_{m,a}\mathbf v_a .
    \label{eq:mic-total-source-mech-map}
\end{align}
Charge and mass occupy the same places, and one velocity enters both the current and the momentum. The sources of the field can therefore be written in mechanical quantities, as the mass and momentum densities of the species, weighted by their charge-to-mass ratios:
\begin{align}
    \rho_{\rm mic}
    &= \sum_a\frac{q_a}{m_a}\rho_{m,a},
    & \mathbf j_{\rm mic}
    &= \sum_a\frac{q_a}{m_a}
       \mathbf g_{{\rm mech},a}^{\rm mic}.
    \label{eq:mic-sources-as-mechanical-projections}
\end{align}

The microscopic fields $\mathbf e$ and $\mathbf b$ obey Maxwell's equations with these sources,
\begin{align}
    \nabla\cdot\mathbf e
    &= \frac{\rho_{\rm mic}}{\varepsilon_0},
    & \nabla\times\mathbf e
    &= -\frac{\partial\mathbf b}{\partial t},
    \nonumber\\
    \nabla\cdot\mathbf b &= 0,
    & \nabla\times\mathbf b
    &= \mu_0\mathbf j_{\rm mic}
       +\mu_0\varepsilon_0\frac{\partial\mathbf e}{\partial t}.
    \label{eq:mic-maxwell-vacuum-form}
\end{align}
All charge and current enter explicitly, and $\varepsilon_0$ and $\mu_0$ are the only field constants; we call this the \emph{vacuum form}~\cite{Jackson1999}. Each carrier has a smooth, localized distribution of charge and mass of small finite extent, so that all products of the fields are finite.

Read in the other direction, the same equations give the sources in terms of the fields:
\begin{align}
    \rho_{\rm mic}
    &= \varepsilon_0\nabla\cdot\mathbf e,
    & \mathbf j_{\rm mic}
    &= \frac{1}{\mu_0}\nabla\times\mathbf b
       -\varepsilon_0\frac{\partial\mathbf e}{\partial t}.
    \label{eq:mic-sources-from-fields}
\end{align}

\emph{The fields say where charged mass is and how it moves.} By Eqs.~\eqref{eq:mic-sources-as-mechanical-projections} and \eqref{eq:mic-sources-from-fields}, charged mass is present wherever $\nabla\cdot\mathbf e$ is not zero, and charged mass moves wherever $\nabla\times\mathbf b/\mu_0-\varepsilon_0\partial_t\mathbf e$ is not zero. The fields give the totals. Where carriers of different species overlap, the totals do not separate them; their individual masses and velocities belong to the description of the carriers.

\subsection{The momentum balance is the force on this mass}
\label{subsec:mic-momentum-equation}

Newton's law for each species takes the local form
\begin{equation}
    \frac{\partial\mathbf g_{{\rm mech},a}^{\rm mic}}{\partial t}
    +\nabla\cdot\mathbf t_{{\rm kin},a}^{\rm mic}
    = \mathbf f_a^{\rm L}+\mathbf f_{{\rm con},a}^{\rm mic}.
    \label{eq:mic-matter-momentum-balance}
\end{equation}
Here $\mathbf t_{{\rm kin},a}^{\rm mic}=\rho_{m,a}\mathbf v_a\otimes\mathbf v_a$ is the momentum flux carried by the moving carriers, with one velocity per species at each point. The left side is the rate of change of the momentum of species $a$, including its transport.

Two forces act. The first is the Lorentz force of the field. It acts where the charge of the species is, and therefore where its mass is:
\begin{equation}
    \mathbf f_a^{\rm L}
    = \rho_a\mathbf e+\mathbf j_a\times\mathbf b .
    \label{eq:species-lorentz-force}
\end{equation}
The second, $\mathbf f_{{\rm con},a}^{\rm mic}$, stands for everything else that acts on the carriers. It represents the stabilization of the carriers and of the material configurations they form, including quantum effects, which this model does not resolve. We call it the constraint force.

Summing the Lorentz force over species gives
\begin{equation}
    \mathbf f_{\rm L}^{\rm mic}
    = \rho_{\rm mic}\mathbf e
      +\mathbf j_{\rm mic}\times\mathbf b .
    \label{eq:mic-lorentz-force-total}
\end{equation}
Inserting the sources from Eq.~\eqref{eq:mic-sources-from-fields} and using Maxwell's equations gives the exact identity~\cite{Jackson1999,FanoChuAdler1960}
\begin{equation}
    \mathbf f_{\rm L}^{\rm mic}
    = \nabla\cdot\mathbf T_{\rm vac}^{\rm mic}
      -\frac{\partial\mathbf g_{\rm vac}^{\rm mic}}{\partial t},
    \label{eq:mic-vac-force}
\end{equation}
where
\begin{align}
    \mathbf g_{\rm vac}^{\rm mic}
    &= \varepsilon_0\mathbf e\times\mathbf b,
    \nonumber\\
    \mathbf T_{\rm vac}^{\rm mic}
    &= \varepsilon_0\left(
       \mathbf e\otimes\mathbf e-\frac12|\mathbf e|^2\mathbf I
       \right)
    \nonumber\\
    &\quad+\frac{1}{\mu_0}\left(
       \mathbf b\otimes\mathbf b-\frac12|\mathbf b|^2\mathbf I
       \right)
    \label{eq:mic-vac-momentum-stress}
\end{align}
are the momentum density of the field and the Maxwell stress; the momentum flux of the field is $-\mathbf T_{\rm vac}^{\rm mic}$. Summing the species balances, Eq.~\eqref{eq:mic-matter-momentum-balance}, then gives
\begin{equation}
\begin{aligned}
    \left(
       \frac{\partial\mathbf g_{\rm mech}^{\rm mic}}{\partial t}
       +\nabla\cdot\mathbf t_{\rm kin}^{\rm mic}
    \right)-\mathbf f_{\rm con}^{\rm mic}
    &= \mathbf f_{\rm L}^{\rm mic}
    \\
    &= \nabla\cdot\mathbf T_{\rm vac}^{\rm mic}
       -\frac{\partial\mathbf g_{\rm vac}^{\rm mic}}{\partial t},
\end{aligned}
    \label{eq:mic-closed-momentum-ledger}
\end{equation}
where $\mathbf t_{\rm kin}^{\rm mic}=\sum_a\mathbf t_{{\rm kin},a}^{\rm mic}$ and $\mathbf f_{\rm con}^{\rm mic}=\sum_a\mathbf f_{{\rm con},a}^{\rm mic}$.

\emph{The momentum balance of the field is the force on charged mass.} The last expression in Eq.~\eqref{eq:mic-closed-momentum-ledger} contains only $\mathbf e$ and $\mathbf b$, and it equals the Lorentz force on the carriers that enter Newton's law on the left. Wherever this field expression is not zero, charged mass is present at that place and receives this force. In a region without carriers the expression vanishes: the stress divergence and the rate of change of the field momentum may each be nonzero there, but they cancel, and momentum is transported through the field without being given to matter. Only the complete expression gives the force; neither of its two terms does.

The converse does not hold. Where the expression is zero, matter may still be present: the forces on individual carriers may vanish or cancel. What balances the force is read from the same equation: the inertia of the carriers and the constraint forces, which hold the loaded carriers in place or pass the load on to other matter.

\subsection{The energy balance is the work of the same force}
\label{subsec:mic-energy-projection}

The energy balance follows from the same equations of motion. Take the scalar product of each species balance, Eq.~\eqref{eq:mic-matter-momentum-balance}, with the velocity of \emph{that species}, and then sum. With the continuity of mass for each species this gives
\begin{equation}
    \frac{\partial u_{\rm mat}^{\rm mic}}{\partial t}
    +\nabla\cdot\mathbf s_{\rm mat}^{\rm mic}
    = \sum_a\mathbf f_a^{\rm L}\cdot\mathbf v_a
      +\sum_a\mathbf f_{{\rm con},a}^{\rm mic}\cdot\mathbf v_a,
    \label{eq:mic-carrier-energy-balance}
\end{equation}
where
\begin{equation}
\begin{aligned}
    u_{\rm mat}^{\rm mic}
    &=\frac12\sum_a\rho_{m,a}|\mathbf v_a|^2,
    \\
    \mathbf s_{\rm mat}^{\rm mic}
    &=\frac12\sum_a\rho_{m,a}|\mathbf v_a|^2\mathbf v_a
\end{aligned}
    \label{eq:mic-carrier-kinetic-moments}
\end{equation}
are the kinetic energy density of the carriers and its flux.

The constraint forces are required to do no net work: their total power vanishes at every instant for the complete assembly and its admissible motions,
\begin{equation}
\begin{aligned}
    \mathcal P_{\rm con}^{\rm mic}
    &\equiv\sum_a\mathbf f_{{\rm con},a}^{\rm mic}\cdot\mathbf v_a,\\
    \int_{\Omega_{\rm complete}}
        \mathcal P_{\rm con}^{\rm mic}(\mathbf x,t)\,dV&=0.
\end{aligned}
    \label{eq:mic-constraint-workless}
\end{equation}
Here $\Omega_{\rm complete}$ contains all constrained carriers and their supports. The constraint forces may pass energy from one carrier or place to another, but they store none. The kinetic energy of the carriers and the energy of the microscopic field are then the only two stores.

The power of the Lorentz force on species $a$ is
\begin{align}
    P_a
    &= \mathbf f_a^{\rm L}\cdot\mathbf v_a
    \nonumber\\
    &= \rho_a\mathbf e\cdot\mathbf v_a
       +(\mathbf j_a\times\mathbf b)\cdot\mathbf v_a
     = \mathbf j_a\cdot\mathbf e,
    \label{eq:species-lorentz-power}
\end{align}
because the magnetic force is perpendicular to the velocity of the charge it acts on:
\begin{equation}
    (\mathbf j_a\times\mathbf b)\cdot\mathbf v_a
    =\rho_a(\mathbf v_a\times\mathbf b)\cdot\mathbf v_a=0 .
    \label{eq:mic-magnetic-work-zero}
\end{equation}
The velocity in $\mathbf j_a=\rho_a\mathbf v_a$ is the velocity of the charged mass that receives the force. The magnetic force can therefore turn the momentum of a carrier, but it cannot change its kinetic energy.

Summing over species gives
\begin{equation}
    P_{\rm L}^{\rm mic}
    =\sum_a\mathbf f_a^{\rm L}\cdot\mathbf v_a
    =\mathbf j_{\rm mic}\cdot\mathbf e .
    \label{eq:mic-total-force-power}
\end{equation}
The order of the two steps matters. If the forces are summed first, the information needed for the power is lost, and no single velocity can stand in for the velocities of the individual carriers. For two equal and opposite forces,
\begin{equation}
    \mathbf f_1+\mathbf f_2=\mathbf0,
    \qquad
    P_1+P_2=\mathbf f_1\cdot(\mathbf v_1-\mathbf v_2),
    \label{eq:mic-opposite-force-work}
\end{equation}
so that zero net force is compatible with nonzero power. Conversely, the magnetic force transfers momentum at zero power. Force and power are connected only through the pairing of each force with the velocity of the mass it acts on.

Inserting the current from Eq.~\eqref{eq:mic-sources-from-fields} into Eq.~\eqref{eq:mic-total-force-power} and using Faraday's law gives Poynting's theorem for this power. With the energy density and the energy flux of the field,
\begin{equation}
\begin{aligned}
    u_{\rm vac}^{\rm mic}
    &=\frac{\varepsilon_0}{2}|\mathbf e|^2
      +\frac{1}{2\mu_0}|\mathbf b|^2,
    \\
    \mathbf S_{\rm vac}^{\rm mic}
    &=\frac{1}{\mu_0}\mathbf e\times\mathbf b,
\end{aligned}
    \label{eq:mic-vac-energy}
\end{equation}
the field identity and the carrier balance read~\cite{Jackson1999,FanoChuAdler1960}
\begin{align}
    -\left(
       \frac{\partial u_{\rm vac}^{\rm mic}}{\partial t}
       +\nabla\cdot\mathbf S_{\rm vac}^{\rm mic}
       \right)
    &=\mathbf j_{\rm mic}\cdot\mathbf e
     =\sum_a\mathbf f_a^{\rm L}\cdot\mathbf v_a
    \nonumber\\
    &=\frac{\partial u_{\rm mat}^{\rm mic}}{\partial t}
      +\nabla\cdot\mathbf s_{\rm mat}^{\rm mic}
      -\mathcal P_{\rm con}^{\rm mic}.
    \label{eq:mic-two-sided-energy-ledger}
\end{align}

\emph{The energy balance of the field is the work of that same force, at the velocity of the mass that receives it.} The first expression in Eq.~\eqref{eq:mic-two-sided-energy-ledger} contains only $\mathbf e$ and $\mathbf b$, and it equals the rate at which the Lorentz forces of the momentum balance do work on the carriers. Wherever it is not zero, at least one carrier is present at that place and moves with a velocity component along the Lorentz force on it. Positive power passes energy from the field to the carriers; negative power returns it. In a region without carriers the expression vanishes: the field energy at a point may change there, but only by transport. As with the force, only the complete expression gives the power.

Here too the converse does not hold. Zero power is compatible with carriers at rest, with forces perpendicular to the motion, and with positive and negative work that cancel. Where the work goes is read from the last line of Eq.~\eqref{eq:mic-two-sided-energy-ledger}: into the kinetic energy of the carriers and, through the constraint forces, to other carriers.

\subsection{The four questions, answered}
\label{subsec:fecc}

The two balances belong together. Collecting Eqs.~\eqref{eq:mic-closed-momentum-ledger} and \eqref{eq:mic-two-sided-energy-ledger},
\begin{equation}
\boxed{\begin{aligned}
    \nabla\cdot\mathbf T_{\rm vac}^{\rm mic}
       -\partial_t\mathbf g_{\rm vac}^{\rm mic}
    &=\sum_a\mathbf f_a^{\rm L},
    \\
    -\left(\partial_tu_{\rm vac}^{\rm mic}
       +\nabla\cdot\mathbf S_{\rm vac}^{\rm mic}\right)
    &=\sum_a\mathbf f_a^{\rm L}\cdot\mathbf v_a.
\end{aligned}}
    \label{eq:mic-fecc-general}
\end{equation}
The left sides contain the four entries of the field tensor: the energy density, the energy flux, the momentum density, and the stress. The right sides contain the same forces twice, once summed and once multiplied by the velocity of the carrier that each acts on. The four questions therefore have definite answers:
\begin{enumerate}
\item[(i)] The carriers receive the force, each where its charge is.
\item[(ii)] The work is done at the velocity of the carrier that receives the force; it is the velocity that appears in the current.
\item[(iii)] The force is balanced by the inertia of the carrier and by the constraint forces on it.
\item[(iv)] The work goes into the kinetic energy of the carrier and, through the constraint forces, to other carriers; negative work returns energy to the field.
\end{enumerate}
The first two answers are fixed by the field identities themselves. The last two are given by the equations of motion of the matter.

The four entries form one object. With $x^0=ct$, the energy--momentum tensor of the field is
\begin{equation}
    T_{\rm vac}^{\mu\nu}
    =\begin{pmatrix}
       u_{\rm vac}^{\rm mic} & (\mathbf S_{\rm vac}^{\rm mic})^T/c\\
       c\mathbf g_{\rm vac}^{\rm mic} & -\mathbf T_{\rm vac}^{\rm mic}
      \end{pmatrix},
    \label{eq:mic-vac-four-tensor}
\end{equation}
where $\mathbf S_{\rm vac}^{\rm mic}=c^2\mathbf g_{\rm vac}^{\rm mic}$. In the corresponding relativistic description of the carriers, let $T_{\rm mech}^{\mu\nu}$ contain their energy and momentum, and let $f_{\rm L}^{\mu}=(P_{\rm L}^{\rm mic}/c,\mathbf f_{\rm L}^{\rm mic})$ and $f_{\rm con}^{\mu}$ be the four-force densities of the Lorentz and constraint forces. Then
\begin{equation}
    \partial_\nu T_{\rm mech}^{\mu\nu}-f_{\rm con}^{\mu}
    =f_{\rm L}^{\mu}
    =-\partial_\nu T_{\rm vac}^{\mu\nu}.
    \label{eq:mic-covariant-ledger}
\end{equation}
The spatial components are the force balance, and the temporal component is the power balance divided by $c$: force and work are components of one equation. Subtracting the balance of rest energy and taking the nonrelativistic limit returns the balances used above.

\label{subsec:two-requirements}
The four answers say what it means for four field quantities to be the energy--momentum tensor of a field. We ask the same of any tensor proposed for the electromagnetic field in matter, in two requirements. \emph{Consistency}: its momentum balance is a force on an identified mass, and its energy balance is the work of that force at the velocity of that mass. \emph{Generality}: this holds for every material and every motion, because the field acts on every charge by the same law. Section~\ref{sec:tensor-role} turns to the macroscopic scale and shows that the macroscopic fields $\mathbf E$ and $\mathbf B$ keep this content.

\section{The macroscopic fields and their tensor}
\label{sec:tensor-role}
\label{sec:macroscopic-compression}

This section passes from the carriers and their field to the macroscopic description. Averaging changes which details remain explicit: the fields become smooth, and so do their sources. We show that the averaged charge and current still belong to identified mass, so that the two balances of the averaged fields remain a force on that mass and its work. Everything that the average does not keep is left to the matter; what it is in a given material is the subject of Secs.~\ref{sec:dielectric-fecc-examples} and \ref{sec:moving-matter-continuum-work}. After the three descriptions of the same carriers (Sec.~\ref{subsec:three-descriptions}), three steps follow: the average of the carriers, which keeps charge and mass together (Sec.~\ref{subsec:retention-material-support}), the average of the fields, which obey Maxwell's equations on their own (Sec.~\ref{subsec:field-split}), and what this fixes and what it leaves to the matter (Sec.~\ref{subsec:macroscopic-fecc}).

\subsection{Three descriptions of the same carriers}
\label{subsec:three-descriptions}

The debate is held at the macroscopic scale, where matter is described by three theories: macroscopic electrodynamics, continuum mechanics, and thermodynamics. Microscopically all three describe the same carriers, the ions and the electrons, and their field, $\mathbf e$ and $\mathbf b$ (Sec.~\ref{sec:microscopic-baseline}). No description can follow them one by one, so each theory averages in space and keeps a few smooth variables.

The motion of the carriers is a sum of three parts: a motion common to all carriers in a neighborhood, a motion of the electrons relative to the ions, and a random remainder. Continuum mechanics keeps the common motion, as the velocity and the strain of one body. Macroscopic electrodynamics keeps the ordered relative arrangement and motion of the charges, as net charge and current, and with them the averaged field. Thermodynamics keeps the random remainder, as the temperature~\cite[Sec.~II\,B]{SchulteWesthoffLocalLoads2026}, \cite[Sec.~2.2]{SchulteWesthoffDielectricEnergy2026}. These are three compressed descriptions of the same carriers, and they exchange momentum and energy with one another.

\subsection{From carriers to two continua}
\label{subsec:retention-material-support}

Macroscopic electrodynamics keeps the ordered arrangement and motion of the charges. We first follow what its average does to the carriers themselves. The average is a spatial filter $\langle\cdot\rangle_{\ell_{\rm EM}}$, a normalized, centered kernel $K_{\ell_{\rm EM}}$ of width $\ell_{\rm EM}$, which removes the structure at the scale of the carriers: $d_{\rm mic}\ll\ell_{\rm EM}\ll L_{\rm EM}$, where $d_{\rm mic}$ is the carrier or lattice scale and $L_{\rm EM}$ the smallest scale on which the averaged fields vary~\cite[Sec.~4.1]{SchulteWesthoffDielectricEnergy2026}. The same filter is applied to the mass, charge, momentum, and current of each carrier~\cite[Sec.~3.2]{SchulteWesthoffDielectricEnergy2026}; the averaging is that of Refs.~\cite[Sec.~III\,A]{SchulteWesthoffLocalLoads2026} and \cite[Secs.~3 and 4]{SchulteWesthoffDielectricEnergy2026}.

For one carrier $p$, of mass $m_p$, charge $q_p$, position $\mathbf r_p(t)$, and velocity $\mathbf v_p(t)$, write $\delta_p=\delta(\mathbf x-\mathbf r_p(t))$ and $K_p=K_{\ell_{\rm EM}}(\mathbf x-\mathbf r_p(t))$. The filter replaces its distributions of mass and charge, written as points on the scale of the filter, $\rho_{m,p}^{\rm mic}=m_p\delta_p$ and $\rho_{q,p}^{\rm mic}=q_p\delta_p$, by $\rho_{m,p}^{\rm ret}=m_pK_p$ and $\rho_{q,p}^{\rm ret}=q_pK_p$, and gives the momentum and the current
\begin{equation}
\begin{aligned}
    \mathbf G_p^{\rm ret}
    &=\langle\rho_{m,p}^{\rm mic}\mathbf v_p\rangle_{\ell_{\rm EM}}
      =\rho_{m,p}^{\rm ret}\mathbf v_p,\\
    \mathbf J_p^{\rm ret}
    &=\langle\rho_{q,p}^{\rm mic}\mathbf v_p\rangle_{\ell_{\rm EM}}
      =\rho_{q,p}^{\rm ret}\mathbf v_p.
\end{aligned}
\label{eq:single-carrier-retained-ordered-moments}
\end{equation}
The average spreads the carrier over the width of the filter, but its mass and its charge are spread in the same way and move with the same velocity $\mathbf v_p(t)$.

Summing these moments separately for the ions and for the electrons defines two continua that occupy the same space:
\begin{equation}
\begin{aligned}
    N_a&=\langle n_a\rangle_{\ell_{\rm EM}},
    &N_a\mathbf V_a&=\langle n_a\mathbf v_a\rangle_{\ell_{\rm EM}},\\
    \rho_{a,{\rm ret}}&=q_aN_a,
    &\rho_{m,a}^{\rm ret}&=m_aN_a,\\
    \mathbf J_a&=\rho_{a,{\rm ret}}\mathbf V_a,
    &\mathbf G_a&=\rho_{m,a}^{\rm ret}\mathbf V_a.
\end{aligned}
\label{eq:retained-species-moments}
\end{equation}
Where $N_a>0$, these moments define the mean velocity $\mathbf V_a$ of species $a$, and $\mathbf J_a=(q_a/m_a)\mathbf G_a$: the current of a species is carried by the momentum of its mass, as in Eq.~\eqref{eq:mic-sources-as-mechanical-projections}. The total charge and current of the two continua are the averages of the microscopic ones, $\rho_{\rm ret}=\sum_a\rho_{a,{\rm ret}}=\langle\rho_{\rm mic}\rangle_{\ell_{\rm EM}}$ and $\mathbf J_{\rm total}=\sum_a\mathbf J_a=\langle\mathbf j_{\rm mic}\rangle_{\ell_{\rm EM}}$.

Two velocities are kept, $\mathbf V_i$ for the ions and $\mathbf V_e=\mathbf V_i+\mathbf W$ for the electrons, so that $\mathbf J_{\rm total}=\rho_{\rm ret}\mathbf V_i+\rho_{e,{\rm ret}}\mathbf W$. Motion of both species together carries mass and momentum but, where the matter is neutral, neither charge nor current; the relative motion $\mathbf W$ carries the current of polarization and conduction. The deviations of the individual carrier velocities from $\mathbf V_a$ carry no net current of the species by construction. Their momentum transport and their energy enter the thermal and mechanical description of the material~\cite[Secs.~2.2 and 3.2]{SchulteWesthoffDielectricEnergy2026}. Ions that move against each other, as in ionic polarization, are further continua of the same kind, each with its own velocity; the balances that follow are sums over the species and keep their form.

\emph{The average keeps charge and mass together.} The charge and the current of each continuum are those of its mass, moving at its velocity. The single carriers are gone from the description, and two continua stand in their place.

\subsection{One field in two parts}
\label{subsec:field-split}

\begin{figure*}[t]
    \centering
    \includegraphics[width=\textwidth]{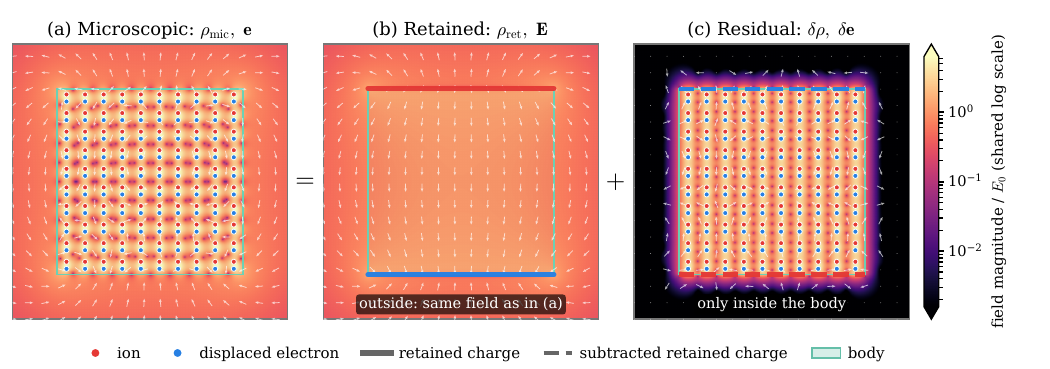}
\caption{One field, two parts, for a polarized dielectric. (a) Microscopic: a $10\times10$ array of ions and displaced electrons, and its field $\mathbf e$. (b) Retained: the smooth charge $\rho_{\rm ret}$, idealized as two sheets at the faces, and its field $\mathbf E$, the field of macroscopic electrodynamics. (c) Residual: $\delta\rho=\rho_{\rm mic}-\rho_{\rm ret}$ and $\delta\mathbf e=\mathbf e-\mathbf E$. Outside the body the fields of (a) and (b) agree; the residual field is confined to the body, where it holds the structure at the scale of the carriers. Color shows the field magnitude on one logarithmic scale, in units of $E_0$, the retained field at the center of the body; arrows show the direction. Two-dimensional model with one regularized kernel for all sources, so that $\mathbf e=\mathbf E+\delta\mathbf e$ holds exactly. The applied field is smooth, belongs to the retained part, and is omitted. Adapted from Fig.~1 of Ref.~\cite{SchulteWesthoffDielectricEnergy2026}.}
    \label{fig:field-two-parts}
\end{figure*}

The fields are averaged with the same filter (Fig.~\ref{fig:field-two-parts}). Sources and fields are thereby divided into two parts,
\begin{equation}
\begin{aligned}
    \rho_{\rm mic}&=\rho_{\rm ret}+\delta\rho,
    &\mathbf j_{\rm mic}&=\mathbf J_{\rm total}+\delta\mathbf j,\\
    \mathbf e&=\mathbf E+\delta\mathbf e,
    &\mathbf b&=\mathbf B+\delta\mathbf b,
\end{aligned}
\label{eq:retained-residual-split}
\end{equation}
of which the first is the average: $\rho_{\rm ret}$ and $\mathbf J_{\rm total}$ are the charge and current of the two continua, $\mathbf E=\langle\mathbf e\rangle_{\ell_{\rm EM}}$, and $\mathbf B=\langle\mathbf b\rangle_{\ell_{\rm EM}}$. We call $\mathbf E$ and $\mathbf B$ the \emph{retained fields}, Fig.~\ref{fig:field-two-parts}(b), and $\delta\mathbf e$ and $\delta\mathbf b$ the \emph{residual fields}, Fig.~\ref{fig:field-two-parts}(c). Away from matter the retained fields are the whole field; inside matter, the residual fields hold what the filter has removed.

The microscopic fields obey Maxwell's equations with all charge and current of the carriers as their sources, Eq.~\eqref{eq:mic-maxwell-vacuum-form}. These equations are linear, and the fixed filter commutes with the derivatives in space and time, so that their average has the same form~\cite[Sec.~6.6]{Jackson1999}. With the fields on the left and the sources on the right,
\begin{equation}
\begin{aligned}
    \varepsilon_0\nabla\cdot\mathbf E&=\rho_{\rm ret},
    &\nabla\cdot\mathbf B&=0,\\
    \frac{1}{\mu_0}\nabla\times\mathbf B-\varepsilon_0\partial_t\mathbf E&=\mathbf J_{\rm total},
    &\nabla\times\mathbf E+\partial_t\mathbf B&=\mathbf 0.
\end{aligned}
\label{eq:retained-maxwell}
\end{equation}
The left sides contain the retained fields and the constants $\varepsilon_0$ and $\mu_0$ alone; the right sides are the charge and current of the two continua. The residual fields obey equations of the same form with the residual sources $\delta\rho$ and $\delta\mathbf j$. The linearity is that of Maxwell's equations for given sources; the response of the material, which determines the sources, may be nonlinear.

\emph{The macroscopic fields are a part of the microscopic field, and they obey Maxwell's equations on their own.} The macroscopic Maxwell equations thus govern a part of the field, not the whole, and no residual quantity enters them. The retained fields are not an approximation to some other field inside matter: they are defined as this part, and they are the fields for which the macroscopic Maxwell equations are solved. The two parts remain connected through the carriers, whose motion generates both sets of sources, and they meet again in every quantity that is quadratic in the fields, such as energy and stress~\cite[Sec.~4.3]{SchulteWesthoffDielectricEnergy2026}.

\emph{The sources of the macroscopic fields are the charge of averaged mass, moving at its velocity.} Where $\nabla\cdot\mathbf E$ is not zero, charged mass is present within the width of the filter; where $\nabla\times\mathbf B/\mu_0-\varepsilon_0\partial_t\mathbf E$ is not zero, charged mass moves. These are the first two rows of Table~\ref{tab:mechanical-content}: the macroscopic Maxwell equations carry the same mechanical information as the microscopic ones, now for two continua in place of single carriers.

Polarization and magnetization are a way of writing part of these sources, and nothing more. With a chosen division of the charge and current into free and bound parts,
\begin{equation}
\begin{aligned}
    \rho_{\rm ret}&=\rho_{\rm free}-\nabla\cdot\mathbf P,\\
    \mathbf J_{\rm total}&=\mathbf J_{\rm free}
       +\frac{\partial\mathbf P}{\partial t}+\nabla\times\mathbf M.
\end{aligned}
\label{eq:source-representation}
\end{equation}
The familiar form of the equations moves these terms to the left, into $\mathbf D=\varepsilon_0\mathbf E+\mathbf P$ and $\mathbf H=\mathbf B/\mu_0-\mathbf M$, which combine the retained fields with variables of the matter. $\mathbf P$ and $\mathbf M$ describe how the bound charge is arranged and moves. They stand for charge and current of the carriers and add no independent electromagnetic degrees of freedom~\cite[Sec.~4.3]{SchulteWesthoffDielectricEnergy2026}. Nor is the representation unique: replacing $\mathbf P$ by $\mathbf P+\nabla\times\mathbf G$ and $\mathbf M$ by $\mathbf M-\partial_t\mathbf G$ leaves the sources and the fields unchanged for any $\mathbf G$. The retained fields are $\mathbf E$ and $\mathbf B$. They are not in dispute; the debate is about what they pass to the matter.

\subsection{What the fields fix, and what remains with the matter}
\label{subsec:macroscopic-fecc}
\label{subsec:retained-electromagnetic-supply-host-complement}
\label{subsec:no-tensor-load}

Sections~\ref{subsec:retention-material-support} and \ref{subsec:field-split} gave the two continua and the fields of their charge and current. It remains to find the forces on the two continua. On each carrier the field acts by the Lorentz force, which is linear in the field, so that it is the sum of two parts,
\begin{equation}
\begin{aligned}
    q_p(\mathbf e+\mathbf v_p\times\mathbf b)
    &=q_p(\mathbf E+\mathbf v_p\times\mathbf B)\\
    &\quad+q_p(\delta\mathbf e+\mathbf v_p\times\delta\mathbf b),
\end{aligned}
\label{eq:carrier-force-two-shares}
\end{equation}
with the fields taken at the carrier. The first part is the force of the retained fields on that carrier, and its work is done at the velocity $\mathbf v_p$ of that carrier. Summed over the carriers of species $a$ and averaged, it is
\begin{equation}
    \mathbf f_{{\rm L},{\rm ret},a}
    =\rho_{a,{\rm ret}}\mathbf E+\mathbf J_a\times\mathbf B
\label{eq:retained-species-lorentz-force}
\end{equation}
to the accuracy of the scale separation; unresolved interfaces are treated in Refs.~\cite{SchulteWesthoffLocalLoads2026,SchulteWesthoffDielectricEnergy2026}. Since $\mathbf J_a=\rho_{a,{\rm ret}}\mathbf V_a$, the magnetic part does no work at the velocity of the species, and $\mathbf f_{{\rm L},{\rm ret},a}\cdot\mathbf V_a=\mathbf J_a\cdot\mathbf E$.

This share is not the whole. Three more things act on the momentum of a species: the residual fields, through the second part of Eq.~\eqref{eq:carrier-force-two-shares}; the constraint forces; and the momentum transport of the carrier motion about the mean. We collect them in one term, the \emph{host force} $\mathbf f_{{\rm host},a}$, and write the momentum balance of species $a$ as
\begin{equation}
    \boldsymbol{\mathcal K}_a^{\rm ret}
      -\mathbf f_{{\rm host},a}
    =\mathbf f_{{\rm L},{\rm ret},a},
\label{eq:retained-host-force-definition}
\end{equation}
where $\boldsymbol{\mathcal K}_a^{\rm ret}=\partial_t\mathbf G_a+\nabla\cdot(\rho_{m,a}^{\rm ret}\mathbf V_a\otimes\mathbf V_a)$ is the rate of change of the momentum of the species, including its transport. This equation defines the host force: it is everything that acts on the species besides the retained fields.

The right side of Eq.~\eqref{eq:retained-host-force-definition} can be written in the fields alone. The retained fields obey Maxwell's equations of vacuum form, Eq.~\eqref{eq:retained-maxwell}, so the algebra of Sec.~\ref{sec:microscopic-baseline} applies to them without change. Summed over the species, their force and its work are the two balances of the vacuum-form tensor of the retained fields:
\begin{equation}
\begin{aligned}
    \mathbf f_{\rm vac}
    &=\sum_a\mathbf f_{{\rm L},{\rm ret},a}
      =\rho_{\rm ret}\mathbf E+\mathbf J_{\rm total}\times\mathbf B\\
    &=\nabla\cdot\mathbf T_{\rm vac}-\partial_t\mathbf g_{\rm vac},\\
    P_{\rm vac}
    &=\sum_a\mathbf f_{{\rm L},{\rm ret},a}\cdot\mathbf V_a
      =\mathbf J_{\rm total}\cdot\mathbf E\\
    &=-\left(\partial_tu_{\rm vac}+\nabla\cdot\mathbf S_{\rm vac}\right).
\end{aligned}
\label{eq:vacuum-force-supply}
\end{equation}
Here $\mathbf T_{\rm vac}$, $\mathbf g_{\rm vac}$, $u_{\rm vac}$, and $\mathbf S_{\rm vac}$ are the expressions of Eqs.~\eqref{eq:mic-vac-momentum-stress} and \eqref{eq:mic-vac-energy}, evaluated with $\mathbf E$ and $\mathbf B$. As relations between the retained fields and their sources, the two identities are exact, and they are known~\cite[Eqs.~(31)--(34)]{Nelson1991}, \cite[Eqs.~(1)--(12)]{ObukhovHehl2003EnergyMomentumForcesMatter}, \cite{KinslerFavaroMcCall2009,Schwarz1992EMForces}. They require no assumption of linear, reversible, nondispersive, or stationary matter. \emph{The momentum balance of the macroscopic fields is their share of the force on the carriers, and their energy balance is the work of that share at the carriers' own velocities.} These are the last two rows of Table~\ref{tab:mechanical-content}.

Summing Eq.~\eqref{eq:retained-host-force-definition} over the two species, and taking its scalar product with the velocity of each species before summing, as in Sec.~\ref{subsec:mic-energy-projection}, gives the pair
\begin{equation}
\boxed{\begin{aligned}
    \boldsymbol{\mathcal K}_{ie}^{\rm ret}-\mathbf f_{\rm host}
       &=\mathbf f_{\rm vac},\\
    \mathcal E_{ie}^{\rm ret}-P_{\rm host}&=P_{\rm vac},
\end{aligned}}
\label{eq:retained-carrier-host-pair}
\end{equation}
with $\boldsymbol{\mathcal K}_{ie}^{\rm ret}=\sum_a\boldsymbol{\mathcal K}_a^{\rm ret}$, $\mathbf f_{\rm host}=\sum_a\mathbf f_{{\rm host},a}$, $\mathcal E_{ie}^{\rm ret}=\sum_a\boldsymbol{\mathcal K}_a^{\rm ret}\cdot\mathbf V_a$, and $P_{\rm host}=\sum_a\mathbf f_{{\rm host},a}\cdot\mathbf V_a$. Here $\mathcal E_{ie}^{\rm ret}$ is the rate of change of the kinetic energy of the two mean motions, including its transport, and the host power $P_{\rm host}$ is the power of the host forces on these motions. On the left stand the two continua and everything else that acts on them; on the right stand the two balances of the tensor, Eq.~\eqref{eq:vacuum-force-supply}.

Equation~\eqref{eq:retained-carrier-host-pair} is not a law of the material. Its right side is fixed by the fields. Its host terms are a name for everything else: they are defined by Eq.~\eqref{eq:retained-host-force-definition}, and no law for them is derived here. Through them the motion that electrodynamics keeps is coupled to the other two descriptions, continuum mechanics and thermodynamics (Sec.~\ref{subsec:three-descriptions}), and what averaging removed from the field returns as a property of the matter, in a dielectric as the binding of the charges (Sec.~\ref{sec:dielectric-fecc-examples}). Predicting them requires the laws and the state of the material.

Energy and momentum divide in the same way. The product $\mathbf J_{\rm total}\cdot\mathbf E$ is the work of the retained fields, not the total work on the carriers~\cite[Sec.~5.2]{SchulteWesthoffDielectricEnergy2026}. What averaging removes from the field has an energy, a momentum, and fluxes of its own, the microscopic expressions minus the retained ones, for example $u_{\rm unres}^{\rm mic}=u_{\rm vac}^{\rm mic}-u_{\rm vac}$. They are not functions of $\mathbf E$ and $\mathbf B$, because they contain the residual fields, in products with the retained fields and in squares. The total energy and momentum of carriers and field, and their fluxes, are therefore sums of three parts:
\begin{equation}
\begin{aligned}
    u_{\rm tot}&=u_{\rm vac}+u_{\rm unres}^{\rm mic}+u_{\rm mat}^{\rm mic},\\
    \mathbf S_{\rm tot}&=\mathbf S_{\rm vac}+\mathbf S_{\rm unres}^{\rm mic}
       +\mathbf s_{\rm mat}^{\rm mic},\\
    \mathbf g_{\rm tot}&=\mathbf g_{\rm vac}+\mathbf g_{\rm unres}^{\rm mic}
       +\mathbf g_{\rm mech}^{\rm mic},\\
    \mathbf T_{\rm tot}&=\mathbf T_{\rm vac}+\mathbf T_{\rm unres}^{\rm mic}
       -\mathbf t_{\rm kin}^{\rm mic}.
\end{aligned}
\label{eq:complete-account}
\end{equation}
The first part belongs to the retained fields, the second to the field structure that averaging removes, and the third to the carriers, Eqs.~\eqref{eq:mic-total-source-mech-map} and \eqref{eq:mic-carrier-kinetic-moments}. For a given averaging the decomposition is exact. The constraint forces store nothing (Sec.~\ref{subsec:mic-energy-projection}); what they pass from one carrier to another is not among these fluxes. The right side of Eq.~\eqref{eq:retained-carrier-host-pair} is what the first part exchanges with the two continua; the second part and the motion of the carriers about the mean act on them through the host terms.

Of the four questions of Sec.~\ref{sec:microscopic-baseline}, this section answers the first two, and it answers them for every material: the ions and the electrons receive the force of the macroscopic fields, as two continua, and its work is done at the velocity of each, $\mathbf V_i$ and $\mathbf V_e$. The other two are open. What balances the force and where the work goes lie in the host terms, and for the same retained fields different materials balance the force and divide the work differently~\cite{SchulteWesthoffLocalLoads2026,SchulteWesthoffDielectricEnergy2026}. The following sections answer them where the material is specified: for a dielectric at rest (Sec.~\ref{sec:dielectric-fecc-examples}) and for a moving and deforming body (Sec.~\ref{sec:moving-matter-continuum-work}).

\section{The dielectric at rest: binding energy and heat}
\label{sec:dielectric-fecc-examples}

We begin with the simplest material: a dielectric whose ions are held at rest, so that only the bound electrons move. The host forces are then the binding of the electrons and, with losses, a drag. This answers the last two questions for this material, what balances the force and where the work goes, before motion and deformation of the body are admitted in Sec.~\ref{sec:moving-matter-continuum-work}.

\subsection{Stationary rigid reversible dielectric}
\label{subsec:rigid-reversible-dielectric-fecc}
\label{subsec:rigid-linear-dielectric-fecc}

Consider an insulating dielectric whose ions form a rigid body at rest, $\mathbf V_i=\mathbf0$, while the bound electrons can move relative to them, $\mathbf V_e=\mathbf W$. There is no free current and no magnetization current, so the current is that of the electrons moving relative to the ions:
\begin{equation}
    \mathbf J_{\rm total}
    =
    \mathbf J_{\rm pol}
    =
    \frac{\partial\mathbf P}{\partial t}
    =
    \sum_{a=i,e}
    \rho_{a,{\rm ret}}\mathbf V_a .
    \label{eq:rigid-reversible-pol-current-paper04}
\end{equation}
By Eq.~\eqref{eq:retained-species-moments}, this current is carried by the mass of the electrons, moving at $\mathbf W$.

The response is taken to be slow and reversible. Exchange with thermal motion is excluded, the inertia and the kinetic energy of the relative motion are negligible, and no energy is radiated. We also neglect transport of unresolved field energy between neighboring material elements and net constraint power within an element. These are the local-storage conditions of Ref.~\cite[Sec.~5.2]{SchulteWesthoffDielectricEnergy2026}.

Under these conditions the host force of Eq.~\eqref{eq:retained-host-force-definition} is reversible. We denote it by $\mathbf f_{{\rm bind},a}$ and call it the binding force. It includes the restoring action of the residual fields on the displaced electrons and the static reaction that holds the ions in place. The momentum balance of each species is
\begin{align}
    \boldsymbol{\mathcal K}_a^{\rm ret}
    -
    \mathbf f_{{\rm bind},a}
    &=
    \rho_{a,{\rm ret}}\mathbf E
    +
    \mathbf J_a\times\mathbf B .
    \label{eq:rigid-reversible-carrier-momentum-paper04}
\end{align}
For slow changes $\boldsymbol{\mathcal K}_a^{\rm ret}\simeq\mathbf0$, so that $\mathbf f_{{\rm bind},a}\simeq-\mathbf f_{{\rm L},{\rm ret},a}$: on each species the binding force balances the Lorentz force of the retained fields.

The two forces on the electrons are opposite, and both act on mass that moves at $\mathbf W$. The work that the Lorentz force does on the electrons is therefore taken up by the binding force at the same velocity, and the electrons gain no appreciable kinetic energy. By Eq.~\eqref{eq:complete-account} the total energy has three parts. The energy has left the first, the retained fields, and the third, the kinetic energy of the carriers, does not change. It is therefore in the second, the field structure that averaging removes: the residual field of Fig.~\ref{fig:field-two-parts}(c). The electrons are sources of both parts of the field and pass the energy from one to the other. Under the local-storage conditions it is stored there recoverably, and it is represented by the binding energy density $u_{\rm bind}$, measured from a chosen reference state of the material~\cite[Sec.~5.2]{SchulteWesthoffDielectricEnergy2026}.

Summing Eq.~\eqref{eq:rigid-reversible-carrier-momentum-paper04} over the two species gives the momentum balance, and taking its scalar product with the velocity of each species before summing gives the energy balance. In the field form of Eq.~\eqref{eq:vacuum-force-supply}, the two balances read, for this dielectric,
\begin{equation}
\begin{aligned}
    \nabla\cdot\mathbf T_{\rm vac}-\partial_t\mathbf g_{\rm vac}
    &=\sum_{a=i,e}\mathbf f_{{\rm L},{\rm ret},a}
     \simeq-\sum_{a=i,e}\mathbf f_{{\rm bind},a},\\
    -\left(\partial_tu_{\rm vac}+\nabla\cdot\mathbf S_{\rm vac}\right)
    &=\mathbf f_{{\rm L},{\rm ret},e}\cdot\mathbf W
     =\mathbf E\cdot\partial_t\mathbf P
     \simeq-\mathbf f_{{\rm bind},e}\cdot\mathbf W
     =\partial_tu_{\rm bind}.
\end{aligned}
\label{eq:rigid-reversible-fecc-pair}
\end{equation}

The first line is a force on the ions and on the electrons. The ions are at rest, so the force on them does no work. The second line is the work of the force on the electrons, at their velocity $\mathbf W$. Positive power increases the binding energy; negative power returns energy to the retained fields. Where the power vanishes, the energy of the retained fields at a point changes only by transport. That the work appears as a rate of change of the binding energy rests on the local-storage conditions. That it is done on the electrons, at $\mathbf W$, does not.

For linear response the binding energy increases by $\tfrac12\mathbf E\cdot\mathbf P$ from the unpolarized state, so that the familiar density $\tfrac12\mathbf E\cdot\mathbf D$ is the electric energy of the retained fields, $\tfrac12\varepsilon_0|\mathbf E|^2$, plus the binding energy. For other reversible responses the stored energy is $\int\mathbf E\cdot d\mathbf P$ along the response curve of the material~\cite[Sec.~6.2]{SchulteWesthoffDielectricEnergy2026}.

A wave of the retained fields carries the binding energy along with its own energy: in a linear, nondispersive dielectric of index $n$, the energy flux $\mathbf S_{\rm vac}$ divided by the energy density of the retained fields, electric and magnetic, is $2nc/(1+n^2)$, which is not the velocity of the wave, and with the binding energy added it is $c/n$. The retained fields pass energy to the binding and take it back within the wave, and the energy travels at the ray velocity~\cite[Sec.~76]{Moller1952} only when both are counted.

\subsection{Stationary rigid lossy dielectric: Binding plus drag}
\label{subsec:lossy-rigid-dielectric-fecc}

We now admit losses, with the ions still at rest and with the same neglect of relative inertia and of transport of unresolved field energy. Exchange between the relative motion and thermal motion is represented by a drag force $\mathbf f_{{\rm drag},a}$ on each species, which acts in addition to the binding force:
\begin{align}
    \boldsymbol{\mathcal K}_a^{\rm ret}
    -
    \mathbf f_{{\rm bind},a}
    -
    \mathbf f_{{\rm drag},a}
    &=
    \rho_{a,{\rm ret}}\mathbf E
    +
    \mathbf J_a\times\mathbf B .
    \label{eq:lossy-carrier-momentum-ledger}
\end{align}
For slow changes, $\mathbf f_{{\rm bind},a}+\mathbf f_{{\rm drag},a}\simeq-\mathbf f_{{\rm L},{\rm ret},a}$. The material laws determine these two forces and the relative current that results.

Binding and drag act on the same moving electrons as the Lorentz force, so the work of all three is done at $\mathbf W$. The binding force stores energy recoverably or returns it. The drag force opposes the relative motion and passes energy to thermal motion, at the rate $q_{\rm th}=-\sum_a\mathbf f_{{\rm drag},a}\cdot\mathbf V_a\ge0$ per unit volume. The two balances of Eq.~\eqref{eq:rigid-reversible-fecc-pair} become
\begin{equation}
\begin{aligned}
    \nabla\cdot\mathbf T_{\rm vac}-\partial_t\mathbf g_{\rm vac}
    &=\sum_{a=i,e}\mathbf f_{{\rm L},{\rm ret},a}
     \simeq-\sum_{a=i,e}\left(\mathbf f_{{\rm bind},a}+\mathbf f_{{\rm drag},a}\right),\\
    -\left(\partial_tu_{\rm vac}+\nabla\cdot\mathbf S_{\rm vac}\right)
    &=\mathbf f_{{\rm L},{\rm ret},e}\cdot\mathbf W
     =\mathbf E\cdot\partial_t\mathbf P
     \simeq-\left(\mathbf f_{{\rm bind},e}+\mathbf f_{{\rm drag},e}\right)\cdot\mathbf W
     =\partial_tu_{\rm bind}+q_{\rm th}.
\end{aligned}
\label{eq:lossy-power-split}
\end{equation}

The energy that the drag removes from the relative motion enters the unordered motion of ions and electrons and the field fluctuations that go with it. Its further storage and transport belong to the thermal balance.

The retained fields fix the sum $\partial_tu_{\rm bind}+q_{\rm th}$: it is the work of their Lorentz force on the electrons. They do not fix how this sum divides into recoverable storage and heat. The binding and drag laws, together with the initial state of the material, determine the division~\cite[Secs.~5.3 and 6.2]{SchulteWesthoffDielectricEnergy2026}.

For this dielectric, the four questions of Sec.~\ref{sec:microscopic-baseline} have the following answers:
\begin{enumerate}
\item[(i)] The ions and the bound electrons receive the Lorentz force of the retained fields.
\item[(ii)] The ions are at rest and receive force without work. The electrons move at $\mathbf W$, and the work on them is $\mathbf E\cdot\partial_t\mathbf P$ per unit volume and time.
\item[(iii)] On each species the force is balanced by binding and drag. For the ions, which are at rest, the binding force includes the reaction that holds them in place.
\item[(iv)] The work goes into the binding energy and into heat.
\end{enumerate}
The first two answers hold for any response of the dielectric, linear or nonlinear, reversible or lossy. The last two depend on the binding and drag laws. We next admit motion and deformation of the body.

\section{Electromagnetic exchange in moving and deforming matter}
\label{sec:moving-matter-continuum-work}

So far the body has been at rest. When it moves and deforms, three things are to be shown. Motion changes how the sources are written, but not which masses receive the force or at which velocity the work is done, and the tensor keeps its form in every inertial frame. The work can now also go into the motion and deformation of the body, and the retained fields fix the sum of what goes into storage, heat, and motion. The retained fields fix the total force and torque on a body. The local load that deforms it follows from the force and the power of the tensor together, once the law of the material for the stored energy is given.

\subsection{Moving sources and covariant force--work balance}
\label{subsec:moving-source-geometry}
\label{subsec:moving-dielectric-source-test}

The retained Maxwell equations have the total charge and current of the carriers as their sources, and polarization and magnetization are one way of writing part of them, Eq.~\eqref{eq:source-representation}. This way of writing depends on the frame. Let $\mathbf P_0$ and $\mathbf M_0$ be the polarization and magnetization in the rest frame of the material. In a frame in which the material moves with velocity $\mathbf V$, the two mix. To leading order in $|\mathbf V|/c$,
\begin{equation}
\begin{aligned}
    \mathbf P'
    &\simeq
    \mathbf P_0
    +
    \frac{1}{c^2}
    \mathbf V\times\mathbf M_0,
    \\
    \mathbf M'
    &\simeq
    \mathbf M_0
    -
    \mathbf V\times\mathbf P_0 ,
\end{aligned}
    \label{eq:low-velocity-source-mixing}
\end{equation}
and the bound sources in that frame are $\rho_{\rm b}'=-\nabla\cdot\mathbf P'$ and $\mathbf J_{\rm b}'=\partial_t\mathbf P'+\nabla\times\mathbf M'$. The following example shows that the two current terms together give the current of the charges that actually move.

\begin{figure}[t]
    \centering
    \includegraphics[width=\onecolfigwidth]{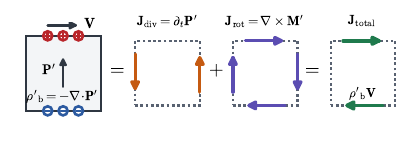}
\caption{One moving source system. The charged top and bottom surfaces carry $\mathbf J_{\rm total}=\rho_{\rm b}'\mathbf V$ (right). The representation terms $\mathbf J_{\rm div}=\partial_t\mathbf P'$ and $\mathbf J_{\rm rot}=\nabla\times\mathbf M'$ (middle) sum to this same current. Their vertical side contributions cancel, consistent with the purely horizontal motion of the surface charges.}
    \label{fig:moving-source-alias}
\end{figure}

Figure~\ref{fig:moving-source-alias} shows a uniformly polarized dielectric that translates rigidly with uniform velocity $\mathbf V=V\hat{\mathbf x}$. Its polarization is fixed in the material and points along $\hat{\mathbf y}$; there are no free sources, and $\mathbf M_0=\mathbf0$. The positive charge on the top surface and the negative charge on the bottom surface move with the body and give opposite convection currents. The uncharged vertical sides carry no net current. Magnetic response in the rest frame, orbital or spin, is left to separate work; magnetic fields and magnetic Lorentz forces are kept.

To leading order in $|\mathbf V|/c$, Eq.~\eqref{eq:low-velocity-source-mixing} gives $\mathbf M'\simeq\mathbf P'\times\mathbf V$, and rigid translation of the fixed polarization pattern gives $\partial_t\mathbf P'=-(\mathbf V\cdot\nabla)\mathbf P'$. At this accuracy,
\begin{align}
    \mathbf J_{\rm b}'
    &=
    \frac{\partial\mathbf P'}{\partial t}
    +
    \nabla\times
    (\mathbf P'\times\mathbf V)
    \nonumber\\
    &=
    -
    \mathbf V(\nabla\cdot\mathbf P')
    =
    \rho_{\rm b}'\mathbf V .
    \label{eq:motional-current-cancellation}
\end{align}
The middle panels of Fig.~\ref{fig:moving-source-alias} show the two terms separately. On the vertical sides they are equal and opposite and cancel; what remains is the horizontal convection current of the charged surfaces. The sum is the current of the charges that move. Neither term alone corresponds to a moving mass or to a mass that receives a force: the magnetization of the moving body is the same charge, written differently.

This example is the low-velocity limit of the covariant description. The bound four-current is
\begin{equation}
    J_{\rm b}^{\nu}
    =
    \partial_\mu M^{\mu\nu},
    \label{eq:bound-four-current-moving}
\end{equation}
where the antisymmetric tensor $M^{\mu\nu}$ contains polarization and magnetization as its frame-dependent parts. Under a Lorentz transformation, this tensor and the four-current transform together. Covariance here concerns the transformation of the retained description once it is formed; it does not assume that an equal-time spatial filter, applied separately in each frame, commutes with a Lorentz boost.
The force and the power of the retained fields, Eq.~\eqref{eq:vacuum-force-supply}, have the same form in every inertial frame, and they remain the force on the two species and its work at the velocity of each (Sec.~\ref{subsec:macroscopic-fecc}). The field tensor $F^{\mu\nu}$ of the retained fields and the total four-current $J_{\rm ret}^{\mu}=(c\rho_{\rm ret},\mathbf J_{\rm total})$ transform as their microscopic counterparts, so the vacuum-form tensor built from $F^{\mu\nu}$ keeps the symmetric, covariant form of Eq.~\eqref{eq:mic-vac-four-tensor}. Let $T_{\rm mech,ret}^{\mu\nu}$ contain the relativistic energy and momentum of the two mean motions, including rest energy, and write $f_{\rm host}^{\mu}=(P_{\rm host}/c,\mathbf f_{\rm host})$ and $f_{\rm vac}^{\mu}=(P_{\rm vac}/c,\mathbf f_{\rm vac})$. Then
\begin{equation}
\begin{aligned}
    \partial_\nu T_{\rm mech,ret}^{\mu\nu}-f_{\rm host}^{\mu}
    &=f_{\rm vac}^{\mu}\\
    &=F^{\mu\nu}J_{{\rm ret},\nu}\\
    &=-\partial_\nu T_{\rm vac}^{\mu\nu}[F].
\end{aligned}
    \label{eq:moving-covariant-vacuum-supply}
\end{equation}
The spatial components are the force balance and the temporal component is the power balance, for the same species, as in Eq.~\eqref{eq:mic-covariant-ledger}. The host term contains the host forces of Sec.~\ref{subsec:retained-electromagnetic-supply-host-complement}, of which binding and drag in Sec.~\ref{sec:dielectric-fecc-examples} are examples. Subtracting the balance of rest energy and taking the nonrelativistic limit returns Eq.~\eqref{eq:retained-carrier-host-pair}.

Motion thus changes how the sources are written. It does not change which masses receive the force of the retained fields or at which velocity the work is done, and the vacuum-form tensor keeps its symmetry and covariance.

\subsection{Coupled carrier motion and energy exchange}
\label{subsec:force-power-energy-exchange}
\label{subsec:deformation-supply-allocation}

We now let the ions move and deform as a body, with velocity $\mathbf V_i=\mathbf V_H$, while the electrons move at $\mathbf V_e=\mathbf V_H+\mathbf W$. This extends the dielectric of Sec.~\ref{subsec:lossy-rigid-dielectric-fecc}. The momentum balance of each species is Eq.~\eqref{eq:retained-host-force-definition}, and the Lorentz forces sum to $\mathbf f_{\rm vac}$, Eq.~\eqref{eq:vacuum-force-supply}, as before; the host force now contains, besides binding and drag, whatever else acts on the species in the moving and deforming body.

The power follows by taking the scalar product with the velocity of each species before summing, as in Eq.~\eqref{eq:retained-carrier-host-pair}:
\begin{equation}
    P_{\rm vac}
      =\mathbf J_{\rm total}\cdot\mathbf E
      =\mathbf f_{\rm vac}\cdot\mathbf V_H
       +\mathbf f_{{\rm L},{\rm ret},e}\cdot\mathbf W .
\label{eq:power-body-relative}
\end{equation}
The first term is the work of the whole force at the velocity of the body, the second the work of the force on the electrons at their velocity relative to it. The matter divides this power among the electrodynamic, the mechanical, and the thermal domain of Ref.~\cite[Sec.~5.3]{SchulteWesthoffDielectricEnergy2026}; with heat supplied by drag,
\begin{equation}
    \underbrace{\mathcal P_{\rm unres}
                 +\mathcal P_{\rm rel}^{\rm kin}}_{\text{electrodynamic}}
      +\underbrace{P_{\rm cont}}_{\text{mechanical}}
      +\underbrace{q_{\rm th}}_{\text{thermal}}
      \simeq P_{\rm vac}.
\label{eq:coupled-carrier-power-projection}
\end{equation}
The power $\mathcal P_{\rm unres}$ is received by the field structure that averaging removes; where this storage is recoverable, it is the rate of change of the binding energy. $\mathcal P_{\rm rel}^{\rm kin}$ is the kinetic power of the motion of the electrons relative to the ions. $P_{\rm cont}$ is the power that goes into the motion and deformation of the body, kinetic and elastic, and $q_{\rm th}$ is the heat supplied by drag. Each is a net power and includes transfer through constraint forces and transport. Motion and recoverable storage can also return energy to the retained fields.

The retained fields fix $P_{\rm vac}$, and with it the sum of what the three domains receive. A nonzero $P_{\rm vac}$ is work of the Lorentz force on moving ions or electrons. The material laws and the state of the material determine how the sum divides among storage, heat, and motion. The first two of the four questions keep their answers of Sec.~\ref{subsec:retained-electromagnetic-supply-host-complement}, whatever the motion of the body; the last two now include its motion and deformation. To predict the deformation, the load that reaches the body must also be known in space. We turn to this next.

\subsection{Determining the mechanical load}
\label{subsec:mechanical-load-closure}

The power $P_{\rm cont}$ in Eq.~\eqref{eq:coupled-carrier-power-projection} is the energy that goes into the motion and deformation of the body. It does not say where in the body the load acts. We now show what the vacuum-form tensor says about this load, following Refs.~\cite{SchulteWesthoffLocalLoads2026,SchulteWesthoffDielectricEnergy2026}. It gives two things. Its force is the force that the retained fields exert on the charges. Its power records the energy passed to the matter, including the field structure that averaging removes, and the rest of the load follows from this energy and its change under deformation. Conditions are quasistatic. We consider first the total force and torque on a body, then the force density in the bulk and the traction at the boundary.

\smallskip
\noindent\emph{Total force and torque.}
Enclose the body in a fixed volume $\mathcal U$ in an inertial frame, with no other matter inside, and choose the surface $\partial\mathcal U$ in vacuum, away from the body and its surface layers. On this surface the microscopic and the retained fields agree to the accuracy of the averaging~\cite[Sec.~4.1]{SchulteWesthoffDielectricEnergy2026}, and so do their Maxwell stresses. With negligible change of field momentum and angular momentum inside $\mathcal U$, both descriptions give the same total electromagnetic force and torque on the body,
\begin{equation}
\begin{aligned}
    \mathbf F_{\rm mic}^{\rm body}
      &\simeq\oint_{\partial\mathcal U}
         \mathbf T_{\rm vac}^{\rm mic}\cdot\mathbf n\,dA\\
      &\simeq\oint_{\partial\mathcal U}
         \mathbf T_{\rm vac}\cdot\mathbf n\,dA,\\[3pt]
    \boldsymbol{\tau}_{\rm mic}^{\rm body}
      &\simeq\oint_{\partial\mathcal U}
         \mathbf r\times\left(\mathbf T_{\rm vac}\cdot\mathbf n\right)\,dA,
\end{aligned}
\label{eq:deformation-global-resultants}
\end{equation}
where $\mathbf n$ is the outward unit normal and $\mathbf r=\mathbf x-\mathbf x_0$ is measured from a fixed origin $\mathbf x_0$. The torque relation uses the symmetry of the vacuum stress. The retained fields therefore determine the total force and torque on a body without any knowledge of the load inside it. Different loads inside can have the same totals and deform the body differently~\cite{MansuripurZakharianWright2013,Markel2023ElectrostaticForceDensity,SunWangNgZhouChan2015}.

To predict the deformation, continuum mechanics needs the force density in the bulk and the traction at the boundary, as they act on the ionic host at the resolution of the mechanical description. This load has two shares. The force of the tensor is the first, and its power leads to the second.

\smallskip
\noindent\emph{The share of the retained fields.}
Summed over the charges, the retained fields exert the force $\mathbf f_{\rm vac}=\rho_{\rm ret}\mathbf E+\mathbf J_{\rm total}\times\mathbf B$, Eq.~\eqref{eq:vacuum-force-supply}. It is known wherever the retained fields are known, for every material. It is a force on the charges, and two things separate it from the load on the host~\cite{SchulteWesthoffLocalLoads2026}: how the force on the electrons reaches the ions through binding and supports, its transmission~\cite[Sec.~IV]{SchulteWesthoffLocalLoads2026}, and the force of the residual fields on the same charges, Eq.~\eqref{eq:carrier-force-two-shares}, which survives averaging in the bulk and, at interfaces, as boundary tractions~\cite[Sec.~V\,B]{SchulteWesthoffLocalLoads2026}. Both belong to the binding of the charges and to the field structure that averaging removes, and no expression in the retained fields contains them: two lattices of the same line dipoles, with the same transmission, polarization, and retained fields, carry different boundary tractions~\cite[Sec.~V\,C and Appendix~A]{SchulteWesthoffLocalLoads2026}.

\smallskip
\noindent\emph{The share of the removed structure.}
The same structure is reached through the power of the tensor, and for the retained fields this is the only access to it. The power $P_{\rm vac}=\mathbf J_{\rm total}\cdot\mathbf E$ is the rate at which energy leaves the retained fields, and Eq.~\eqref{eq:coupled-carrier-power-projection} lists what receives it: the removed field structure, the relative motion of the electrons, the body, and heat. The retained fields at a given state do not contain this energy; it is known only as the integral of the power along the path of the process, and its stored part is the binding energy $U_{\rm bind}$ (Sec.~\ref{sec:dielectric-fecc-examples}). The work on the ionic host in a virtual deformation is then~\cite[Sec.~6.1]{SchulteWesthoffDielectricEnergy2026}
\begin{equation}
    \delta W_{\rm em}
      \simeq\delta W_{\rm ext,elec}
      -\delta\!\left(U_{\rm vac}+U_{\rm bind}\right),
\label{eq:host-load-energy-variation}
\end{equation}
for reversible quasistatic changes, with zero total virtual work of the constraints and negligible electromagnetic energy leaving through the outer boundary. Here $\delta W_{\rm ext,elec}$ is the work of the electrical sources for the controls imposed, and the energies include the field around the body. The retained fields give $\delta W_{\rm ext,elec}$ and $\delta U_{\rm vac}$. The share of the removed structure is the last term, the change of the binding energy under the deformation.

For a linear dielectric the variation gives the Helmholtz stress~\cite[Sec.~16]{LandauLifshitzVol8}, \cite[Sec.~6.2]{SchulteWesthoffDielectricEnergy2026}. For small strain $u_{ik}$ of a solid that is isotropic when undeformed, with $D_l=\varepsilon_{lm}E_m$,
\begin{equation}
\begin{aligned}
    \sigma_{ik}^{\rm elec}
    &=\varepsilon_0\left(E_iE_k-\tfrac12|\mathbf E|^2\delta_{ik}\right)\\
    &\quad+\tfrac12\left(E_iP_k+E_kP_i\right)
      -\tfrac12\,\mathbf E\cdot\mathbf P\,\delta_{ik}\\
    &\quad-\tfrac12E_lE_m
      \left(\frac{\partial\varepsilon_{lm}}{\partial u_{ik}}\right)_{T,\mathbf E}.
\end{aligned}
\label{eq:helmholtz-stress-three-parts}
\end{equation}
The first line is the electric stress of the retained fields, the stress of the vacuum-form tensor. The other two lines are the stress of the binding energy. The second is fixed by the polarization at the present configuration. The third, the electrostrictive stress, requires the dependence of the permittivity on strain, which differs between materials of equal permittivity~\cite{LeePengShkel2005,SunWangNgZhouChan2015}; it is measured by the same record at neighboring deformations, or computed at finer resolution. Force and power of the one tensor thus lead to the load of the linear dielectric. For nonlinear and for lossy response the route is the same, with the stored part of the path integral in place of $\tfrac12\mathbf E\cdot\mathbf P$; only the information required of the material grows~\cite[Secs.~6.2 and 6.3]{SchulteWesthoffDielectricEnergy2026}.

\emph{The vacuum-form tensor thus holds what the retained fields can say about the load: its force is their share of the force on the charges, and its power is the record from which the share of the removed structure follows.} What must be added is the law of the material: how the stored energy changes with deformation. No expression in the retained fields alone can replace this law, whichever tensor it is taken from, and a comparison of force densities therefore cannot decide between tensors. The constitutive laws of continuum mechanics, for the energy or for the load with a specified transmission, are laws of this kind~\cite{ToupinElasticDielectric,Maugin1980VirtualPower,McMeekingLandis2005StoredEnergyDeformableDielectrics,SuoZhaoGreene2008NonlinearFieldTheory,DorfmannOgdenNonlinearElectroelasticity}.

\section{The map of the three descriptions}
\label{sec:mechanical-unification}

The preceding sections followed the exchange between the retained fields and the matter into storage, heat, motion, and deformation. We now put the parts together, with Refs.~\cite{SchulteWesthoffLocalLoads2026,SchulteWesthoffDielectricEnergy2026}, into one map of electromagnetic multiphysics: what the three descriptions of Sec.~\ref{subsec:three-descriptions} each keep of the same carriers, and how macroscopic electrodynamics is connected to the other two. Each description has its own variables and its own balance law. What averaging removes is not gone from the material: where it matters, a material law has to carry it. The map holds within the classical model of Sec.~\ref{sec:microscopic-baseline} and the scope stated in Sec.~\ref{sec:introduction}.

Table~\ref{tab:multiphysics-architecture} is the map. Its upper block is the field: the retained fields, their vacuum-form tensor, and its two balances. Its lower block is the matter: the motion of the carriers and the field structure that averaging removes, each in its three roles. Macroscopic electrodynamics is the upper block together with the first column: the fields and the electrical response of the material. The energy entries are those of the energy map of Ref.~\cite[Sec.~5.3]{SchulteWesthoffDielectricEnergy2026}, and the entries for the load follow Ref.~\cite{SchulteWesthoffLocalLoads2026}; the present paper adds the tensor, and with it momentum, stress, and the fluxes.

\begin{table*}[t]
\caption{The map of electromagnetic multiphysics. The field holds the first part of Eq.~\eqref{eq:complete-account}; the matter holds the other two, in three roles, each with its own variables and balance law. The force and the power of the retained fields on the two continua are a budget. The roles classify the origin of each entry. They are not by themselves an additive decomposition, since the roles are coupled.}
\label{tab:multiphysics-architecture}
\small
\setlength{\tabcolsep}{5pt}
\renewcommand{\arraystretch}{1.15}
\centering
\begin{tabular}{@{}p{0.17\textwidth}p{0.24\textwidth}p{0.26\textwidth}p{0.23\textwidth}@{}}
\toprule
\raggedright The field
& \multicolumn{3}{p{0.76\textwidth}}{\raggedright Retained fields $\mathbf E,\mathbf B$ of macroscopic electrodynamics, governed by Maxwell's equations} \tabularnewline
\raggedright Tensor
& \multicolumn{3}{p{0.76\textwidth}}{\raggedright Energy $u_{\rm vac}$, flux $\mathbf S_{\rm vac}$, momentum $\mathbf g_{\rm vac}$, and stress $\mathbf T_{\rm vac}$, fixed by $\mathbf E$ and $\mathbf B$ alone} \tabularnewline
\raggedright Budget
& \multicolumn{3}{p{0.76\textwidth}}{\raggedright Force $\mathbf f_{\rm vac}$ on the two continua and its work $P_{\rm vac}$ at their velocities, for every material} \tabularnewline
\midrule
\raggedright The matter
& \raggedright Electrical response
& \raggedright Continuum mechanics
& \raggedright Thermodynamics \tabularnewline
\midrule
\raggedright Role of the carriers; variables
& \raggedright Relative arrangement and motion of charge: $\rho_{\rm ret}$, $\mathbf J_{\rm total}$, $\mathbf W$
& \raggedright Common motion and deformation: velocity $\mathbf V_H$, strain
& \raggedright Unordered motion: temperature \tabularnewline
\raggedright Held by the carriers
& \raggedright Kinetic energy and momentum of the relative motion
& \raggedright Kinetic energy and momentum of the body
& \raggedright Thermal energy, kinetic part of the pressure \tabularnewline
\raggedright Held by the unresolved field structure
& \raggedright Binding energy, stress that passes on the load on the bound charge
& \raggedright Elastic energy, elastic stress
& \raggedright Thermal field energy, its part of the pressure \tabularnewline
\raggedright Balance law
& \raggedright Momentum balance of the relative motion
& \raggedright Momentum balance of the body
& \raggedright Heat balance \tabularnewline
\midrule
\raggedright Input needed
& \raggedright Force of the fields on each continuum
& \raggedright Local load: force density and boundary traction
& \raggedright Heat rate \tabularnewline
\raggedright Fixed by the retained fields
& \raggedright This force, Eq.~\eqref{eq:retained-species-lorentz-force}
& \raggedright Total force and torque on a body; the share $\mathbf f_{\rm vac}$ of the local load
& \raggedright The sum of storage, motion, and heat \tabularnewline
\raggedright Supplied by material laws
& \raggedright Host forces: binding and drag
& \raggedright How the binding energy changes with deformation~\cite{SchulteWesthoffDielectricEnergy2026}
& \raggedright Division of the work~\cite{SchulteWesthoffDielectricEnergy2026} \tabularnewline
\bottomrule
\end{tabular}
\end{table*}

Three statements complete the map, and together they answer the four questions.

\emph{Which mass, and at which velocity: the budget is fixed by the retained fields.} The field exchanges momentum and energy with the matter in one way only: through the force of the retained fields on the two continua, the ions and the electrons, and its work at the velocity of each, Eq.~\eqref{eq:vacuum-force-supply}. No law of the material enters, so this holds for every material and every motion. This force and this work are a budget for the other two descriptions, not their input.

\emph{What balances the force: how the load reaches the body is fixed by the material.} Continuum mechanics needs the local load on the body. The force of the retained fields is the part of it that is known directly; the rest is the change of the binding energy under deformation (Sec.~\ref{subsec:mechanical-load-closure}). The material supplies how this energy depends on deformation~\cite{SchulteWesthoffDielectricEnergy2026} or, with a finer resolution of the material, the rule of transmission and the missing correlations~\cite{SchulteWesthoffLocalLoads2026}. Only the total force and torque on a body are fixed by the retained fields alone.

\emph{Where the work goes: its division is fixed by the material.} Thermodynamics needs the heat rate, which is not the power of the retained fields. This power divides among the energies of the lower block: binding energy and the kinetic energy of the relative motion, the motion and deformation of the body, and heat. Unlike the load, the sum is fixed by the retained fields at every point; the material laws fix the division~\cite{SchulteWesthoffDielectricEnergy2026}.

The coupling of the three descriptions is thus mechanically consistent: every transfer in the map is a force on an identified continuum and its work at the velocity of that continuum, Eq.~\eqref{eq:retained-carrier-host-pair}. The tensor gives the force and the work to which the laws of the material apply; it does not contain these laws.

The map is complete. The total energy and momentum of carriers and field are the sum of three parts, Eq.~\eqref{eq:complete-account}: that of the retained fields, that of the field structure that averaging removes, and that of the carriers. Every part has an owner. The constraint forces hold no energy in this model, but they pass load on, and the table counts that load with the stress of the field structure that they stabilize. For the second and the third part the owners are those of the classical model: a finer model, with energy in the quantum effects that the constraint forces represent, would divide these two parts differently~\cite[Sec.~5.2]{SchulteWesthoffDielectricEnergy2026}. The first part, and the force and the work that it passes on, would be the same. The map gives electromagnetic multiphysics a common mechanical foundation: the fields give the budget, and material laws complete the prediction of storage, heat, motion, and deformation.

The question of the tensor debate is the question of the upper block of this map: which tensor is that of the field.

\section{Testing the other tensors and the prevailing resolutions}
\label{sec:tensor-splits}
\label{sec:audit}

Sections~\ref{sec:microscopic-baseline}--\ref{sec:mechanical-unification} assign every part of the total energy and momentum to the fields or carriers that hold it. We now hold the Minkowski and Abraham tensors to the two requirements of Sec.~\ref{subsec:two-requirements} (Sec.~\ref{subsec:audit-minkowski}). Section~\ref{subsec:audit-protocol} states the result and sets it against the first of the two prevailing resolutions of the debate, which conclude that no single tensor is that of the electromagnetic field in matter: that the division of the total into an electromagnetic and a material part is arbitrary. The second holds that the Abraham and Minkowski momenta are both correct, as the kinetic and the canonical momentum of light (Sec.~\ref{subsec:canonical-kinetic}); it relies on the same freedom of division~\cite{Barnett2010}. For each we separate what it establishes from what is concluded from it. Section~\ref{subsec:canonical-kinetic} begins with what the experiments of the debate measure, a force, and Sec.~\ref{subsec:information-loss-tensor} follows this force to the individual experiments and to the force laws that remain in use.

\subsection{The three tensors in a dielectric}
\label{subsec:audit-minkowski}

\begin{table*}[!t]
\caption{The vacuum-form, Minkowski, and Abraham tensors for a nonmagnetic insulating dielectric at rest, with $\mathbf D=\varepsilon_0\mathbf E+\mathbf P$ and $\mathbf H=\mathbf B/\mu_0$. Upper part: the four entries and the two balances, Eq.~\eqref{eq:residual-audit}. The Minkowski and Abraham entries are the conventional ones~\cite[Eqs.~(40) and (42)]{Pfeifer2007}: $\mathbf T_M=\mathbf E\otimes\mathbf D+\mathbf H\otimes\mathbf B-\tfrac12(\mathbf E\cdot\mathbf D+\mathbf H\cdot\mathbf B)\mathbf I$, and $\mathbf T_A$ is its symmetric part. They add $\Delta\mathbf T_M=\mathbf E\otimes\mathbf P-\tfrac12(\mathbf E\cdot\mathbf P)\mathbf I$ and its symmetric part $\Delta\mathbf T_A$ to the vacuum-form stress; the divergence acts on the second index, $(\nabla\cdot\mathbf T)_i=\partial_jT_{ij}$. Lower part: the answers to the four questions of Sec.~\ref{sec:microscopic-baseline}, and what follows for the heat and for the load on the body. Those of the vacuum-form tensor hold for every response of the dielectric.}
\label{tab:tensor-comparison}
\small
\setlength{\tabcolsep}{5pt}
\renewcommand{\arraystretch}{1.15}
\centering
\begin{tabular}{@{}p{0.18\textwidth}p{0.28\textwidth}p{0.24\textwidth}p{0.19\textwidth}@{}}
\toprule
& \raggedright Vacuum form
& \raggedright Minkowski
& \raggedright Abraham \tabularnewline
\midrule
\raggedright Energy density
& \raggedright $u_{\rm vac}$
& \multicolumn{2}{p{0.45\textwidth}}{\raggedright $\tfrac12(\mathbf E\cdot\mathbf D+\mathbf B\cdot\mathbf H)=u_{\rm vac}+\tfrac12\mathbf E\cdot\mathbf P$ for both} \tabularnewline
\raggedright Energy flux
& \raggedright $\mathbf S_{\rm vac}$
& \multicolumn{2}{p{0.45\textwidth}}{\raggedright $\mathbf E\times\mathbf H=\mathbf S_{\rm vac}$ for both} \tabularnewline
\raggedright Momentum density
& \raggedright $\mathbf g_{\rm vac}$
& \raggedright $\mathbf D\times\mathbf B=\mathbf g_{\rm vac}+\mathbf P\times\mathbf B$
& \raggedright $\mathbf E\times\mathbf H/c^2=\mathbf g_{\rm vac}$ \tabularnewline
\raggedright Stress
& \raggedright $\mathbf T_{\rm vac}$
& \raggedright $\mathbf T_M=\mathbf T_{\rm vac}+\Delta\mathbf T_M$
& \raggedright $\mathbf T_A=\mathbf T_{\rm vac}+\Delta\mathbf T_A$ \tabularnewline
\raggedright Force
& \raggedright $\mathbf f_{\rm vac}=-(\nabla\cdot\mathbf P)\mathbf E+\partial_t\mathbf P\times\mathbf B$
& \raggedright $\mathbf f_{\rm vac}+\nabla\cdot\Delta\mathbf T_M-\partial_t(\mathbf P\times\mathbf B)$
& \raggedright $\mathbf f_{\rm vac}+\nabla\cdot\Delta\mathbf T_A$ \tabularnewline
\raggedright Power
& \raggedright $P_{\rm vac}=\mathbf E\cdot\partial_t\mathbf P$
& \multicolumn{2}{p{0.45\textwidth}}{\raggedright $P_{\rm vac}-\partial_t\!\left(\tfrac12\mathbf E\cdot\mathbf P\right)$ for both} \tabularnewline
\midrule
\raggedright (i) Which mass receives the force?
& \raggedright The ions and the bound electrons, each where its charge is
& \multicolumn{2}{p{0.45\textwidth}}{\raggedright Not the bound charge. The force contains the divergence of an added stress and has been taken to act on the body as a whole~\cite[Chap.~4]{PenfieldHaus1967}} \tabularnewline
\raggedright (ii) At which velocity is the work done?
& \raggedright At the velocity $\mathbf W$ of the bound electrons; the work is $\mathbf E\cdot\partial_t\mathbf P$, Eq.~\eqref{eq:lossy-power-split}. For the moving body, Eq.~\eqref{eq:vacuum-power-deforming}
& \multicolumn{2}{p{0.45\textwidth}}{\raggedright The body is at rest, so the force does no work; the power is not zero for nonlinear or lossy response, Eq.~\eqref{eq:minkowski-power-no-work}. When the linear dielectric deforms, the power contains work that the force does not do, Eq.~\eqref{eq:minkowski-abraham-power-deforming}} \tabularnewline
\raggedright (iii) What balances the force?
& \raggedright Binding and drag on the electrons; the support of the ions
& \multicolumn{2}{p{0.45\textwidth}}{\raggedright The support of the body. Binding and drag do not appear} \tabularnewline
\raggedright (iv) Where does the work go?
& \raggedright Into binding energy and heat
& \multicolumn{2}{p{0.45\textwidth}}{\raggedright There is no work. The power is the delivered energy that the tensor does not hold, Eq.~\eqref{eq:minkowski-abraham-power}} \tabularnewline
\midrule
\raggedright Heat that follows
& \raggedright $P_{\rm vac}$ less the rate of change of the binding energy, at every instant, Eq.~\eqref{eq:lossy-power-split}
& \multicolumn{2}{p{0.45\textwidth}}{\raggedright The power only over a closed cycle, Eq.~\eqref{eq:minkowski-loss-cycle}} \tabularnewline
\raggedright Load on the body that follows
& \raggedright $\mathbf f_{\rm vac}$ and the change of the binding energy under deformation; Eq.~\eqref{eq:helmholtz-stress-three-parts} for linear response
& \multicolumn{2}{p{0.45\textwidth}}{\raggedright One stress for all materials of equal polarization: Eq.~\eqref{eq:helmholtz-stress-three-parts} without its third line} \tabularnewline
\midrule
\raggedright Law of the material contained in the tensor
& \raggedright None
& \multicolumn{2}{p{0.45\textwidth}}{\raggedright The work stored in a linear dielectric, and one rule for passing on the load on the bound charge} \tabularnewline
\bottomrule
\end{tabular}
\end{table*}

\emph{The test.} The test is that of Sec.~\ref{subsec:two-requirements}, put to each tensor in turn. Is its momentum balance a force on an identified mass? Is its energy balance the work of that force, at the velocity of that mass? And does this hold for every response of the material? We also follow what each tensor gives for what is measured: the heat, and the load on the body. A candidate $X$ adds terms to the vacuum-form tensor of the retained fields, $\mathcal T_X=\mathcal T_{\rm vac}+\Delta\mathcal T$ with four-tensor indices suppressed, and so changes the four entries by $(\Delta u,\Delta\mathbf S,\Delta\mathbf g,\Delta\mathbf T)$. Its force and power are $\mathbf f_X=\mathbf f_{\rm vac}+\Delta\mathbf f_X$ and $P_X=P_{\rm vac}+\Delta P_X$, with
\begin{equation}
\begin{aligned}
    \Delta\mathbf f_X
      &=\nabla\cdot\Delta\mathbf T-\partial_t\Delta\mathbf g,\\
    \Delta P_X
      &=-\left(\partial_t\Delta u+\nabla\cdot\Delta\mathbf S\right).
\end{aligned}
\label{eq:residual-audit}
\end{equation}
A candidate that adds an energy must add the stress of that energy as well; if it does not, its two balances no longer fit when the body deforms (Sec.~\ref{subsubsec:audit-linear}).

We take the dielectric of Sec.~\ref{sec:dielectric-fecc-examples}: insulating and nonmagnetic, with no free charge, so that $\rho_{\rm ret}=-\nabla\cdot\mathbf P$. Its ions are at rest, so that $\mathbf J_{\rm total}=\partial_t\mathbf P$; only in Sec.~\ref{subsubsec:audit-linear} is the body also allowed to move and deform slowly. Its response is left open: it may be linear or nonlinear, reversible or lossy. While the body is at rest, nothing moves in it, apart from thermal motion, except the bound electrons, so every exchange of energy between the retained fields and the matter is work on them. Free charge is left out because the three tensors agree on it: each contains the Lorentz force on free charge and the work of that force. Table~\ref{tab:tensor-comparison} sets the three tensors side by side: their four entries, their two balances, and their answers to the four questions. We keep the conventional expressions of the Minkowski and Abraham tensors unchanged for every response; expressions adapted to the response are taken up at the end of Sec.~\ref{subsubsec:audit-nonlinear}.

\emph{Why static and slow fields decide.} The cases that follow are static or slowly varying, and the debate is best known for light. Two questions have to be kept apart. The first is how the fields evolve, and it is not in dispute. Maxwell's equations and the laws of the material fix $\mathbf E$ and $\mathbf B$ together with their sources, the charge and the current, whether these are written as $\rho_{\rm ret}$ and $\mathbf J_{\rm total}$ or as free charge and current with $\mathbf P$ and $\mathbf M$, Eq.~\eqref{eq:source-representation}. They do so for a static field and for a wave alike, down to the velocity of the wave. Every tensor of the debate is written in these same fields and sources, and the choice of a tensor changes none of them. The second question is what the fields pass to the matter: a force, and the work of that force. This is the question of the tensor. The two balances of the vacuum-form tensor are that force and that work for every time dependence, Eq.~\eqref{eq:vacuum-force-supply}. A tensor of the field has to meet the requirements in every process, so one process in which it fails decides, and the simplest shows the failure most plainly. What is measured with light is again a force on matter, not a momentum of the light (Secs.~\ref{subsec:canonical-kinetic} and \ref{subsec:information-loss-tensor}).

The three tensors differ in what they do with the bound charge. The vacuum-form tensor contains $\mathbf P$ only as the charge and the current that it represents, Eq.~\eqref{eq:source-representation}. The other two contain $\mathbf P$ itself, through $\mathbf D$, as if it were a field of its own. The fields and their sources do not fix such a quantity: a static solenoidal field added to $\mathbf P$ changes neither the charge nor the current (Sec.~\ref{subsec:field-split}), but it changes these tensors. What fixes $\mathbf P$ is a statement about the matter, and the failures found below follow from this one step. The vacuum-form tensor passes the test for every response. The other two fail it: in the linear reversible dielectric, their best case, in the load and, when the body deforms, in the match of force and power (Sec.~\ref{subsubsec:audit-linear}); for nonlinear and lossy response even in the energy balance of a body at rest (Sec.~\ref{subsubsec:audit-nonlinear}).

\subsubsection{The linear reversible dielectric}
\label{subsubsec:audit-linear}

This is the case in which the two tensors are usually considered, and the best case for them. For a rigid body in a static field the three tensors cannot be told apart: at fixed deformation the sum of energies is the same, and the added stresses vanish outside the body, so the total force on the body is the same. They differ in the load inside a body that can deform. This load is the Helmholtz stress, Eq.~\eqref{eq:helmholtz-stress-three-parts}~\cite[Sec.~16]{LandauLifshitzVol8}, \cite[Sec.~6.2]{SchulteWesthoffDielectricEnergy2026}. We follow how each tensor accounts for it, one tensor at a time and in the same steps: the energy of the field, what passes to the matter, the stress of the tensor, the load, and the check of the power against the work of the tensor's own force.

\emph{The vacuum-form tensor.} It treats the bound charge as it treats any charge. The energy of the field is $u_{\rm vac}$. The momentum balance is the Lorentz force of the retained fields on the ions and on the bound electrons~\cite{ChuHausPenfield1966,Melcher1981,Reich2017,Rickert2023}, and the energy balance is the work of that force on the electrons, $P_{\rm vac}=\mathbf E\cdot\partial_t\mathbf P$. This work leaves the retained fields. Integrated from the unpolarized state it is $\tfrac12\mathbf E\cdot\mathbf P$~\cite[Sec.~3]{Saldanha2011}, \cite{AnghinoniPartanenAstrath2024LocalModel}: the binding energy, stored in the field structure that averaging removes (Sec.~\ref{subsec:rigid-reversible-dielectric-fecc}) and not in a field $\mathbf P$. The stress of the tensor is the stress of the retained fields, and the change of the binding energy under deformation adds the rest of the load (Sec.~\ref{subsec:mechanical-load-closure}). Together they are the Helmholtz stress:
\begin{equation}
\begin{aligned}
    \sigma_{ik}^{\rm elec}
    &=\underbrace{\varepsilon_0\left(E_iE_k-\tfrac12|\mathbf E|^2\delta_{ik}\right)}_{\text{stress of the tensor}}\\
    &\quad+\underbrace{\tfrac12\left(E_iP_k+E_kP_i\right)
      -\tfrac12\,\mathbf E\cdot\mathbf P\,\delta_{ik}}_{\text{binding energy, fixed by }\mathbf P}\\
    &\quad-\underbrace{\tfrac12E_lE_m
      \left(\frac{\partial\varepsilon_{lm}}{\partial u_{ik}}\right)_{T,\mathbf E}}_{\text{\makebox[0pt]{binding energy, depends on the material}}}.
\end{aligned}
\label{eq:helmholtz-stress-origins}
\end{equation}
The third line is the electrostrictive stress. It depends on how the permittivity changes with strain, which no expression in the retained fields contains; it differs between materials of equal permittivity~\cite{LeePengShkel2005,SunWangNgZhouChan2015}, and it has been measured directly, in dielectric liquids~\cite[Secs.~2.3 and 2.4]{Brevik1979}. The tensor thus gives the first line as its stress and the other two through its power.

A body that moves and deforms also tests a tensor against itself: its power must be the work of its own force (Sec.~\ref{subsec:two-requirements}). For the vacuum-form tensor it is, for every motion, Eq.~\eqref{eq:power-body-relative}:
\begin{equation}
    P_{\rm vac}=\underbrace{\mathbf f_{\rm vac}\cdot\mathbf V_H+\mathbf f_{{\rm L},{\rm ret},e}\cdot\mathbf W}_{\text{work of its own force}},
\label{eq:vacuum-power-deforming}
\end{equation}
at the velocity $\mathbf V_H$ of the body and at the velocity $\mathbf W$ of the electrons relative to it. \emph{For the linear dielectric the force and the power of the vacuum-form tensor thus lead to the whole Helmholtz stress, with the law of the material, and its energy balance contains the work of its own force and nothing else.}

\emph{The Minkowski tensor.} Its energy density is $u_{\rm vac}+\tfrac12\mathbf E\cdot\mathbf P$, with the energy flux of the retained fields. Its power, $P_{\rm vac}-\partial_t(\tfrac12\mathbf E\cdot\mathbf P)$, is zero: the same $\tfrac12\mathbf E\cdot\mathbf P$ is here energy of the field, held by $\mathbf P$, and nothing passes to the matter. The stress of this energy is then the tensor's as well: no stored energy is left in the matter that could load the body, and the stress of the tensor has to be the whole load. In a static electric field, denoted by $(E)$, this stress is (Table~\ref{tab:tensor-comparison})
\begin{equation}
    \left(\mathbf T_M^{(E)}\right)_{ik}
    =\varepsilon_0\left(E_iE_k-\tfrac12|\mathbf E|^2\delta_{ik}\right)
     +E_iP_k-\tfrac12\,\mathbf E\cdot\mathbf P\,\delta_{ik}.
\label{eq:minkowski-static-stress}
\end{equation}
Its symmetric part is the first two lines of Eq.~\eqref{eq:helmholtz-stress-origins}, and for isotropic response it is these two lines themselves. The third line, the electrostrictive stress, is missing~\cite[Sec.~1.2]{Brevik1979}, \cite[Sec.~VI]{Pfeifer2007}: the tensor has taken in the binding energy together with the part of its stress that $\mathbf E$ and $\mathbf P$ fix, and the part that depends on the material has no place in it. With $\mathbf f_{\rm vac}^{(E)}=-(\nabla\cdot\mathbf P)\mathbf E$, the force of the tensor is
\begin{equation}
\begin{aligned}
    \mathbf f_M^{(E)}
    &=\mathbf f_{\rm vac}^{(E)}+\nabla\cdot\Delta\mathbf T_M\\
    &=(\mathbf P\cdot\nabla)\mathbf E
      -\tfrac12\nabla(\mathbf E\cdot\mathbf P).
\end{aligned}
\label{eq:minkowski-linear-electric-force}
\end{equation}
For linear isotropic response with permittivity $\varepsilon$ this is the familiar $-\tfrac12|\mathbf E|^2\nabla\varepsilon$, a force where the permittivity changes. Stress and force are fixed by $\mathbf E$ and $\mathbf P$ alone: they are the same for all materials of equal permittivity. This is not economy. Bodies with the same retained fields and polarization can carry different loads~\cite[Sec.~V\,C]{SchulteWesthoffLocalLoads2026}, \cite{SunWangNgZhouChan2015,WangNgXiaoChan2016}, and the tensor has one.

\emph{The same check for the Minkowski tensor.} For slow motion and deformation in a quasistatic field, of a solid that is isotropic when undeformed and to first order in the strain, with $\mathbf H=\mathbf B/\mu_0-\mathbf M'$ of the moving body, Eq.~\eqref{eq:low-velocity-source-mixing}, its power is~\cite[Eq.~(27)]{Obukhov2008}
\begin{equation}
\begin{aligned}
    P_M&=\tfrac12E_lE_m\,\partial_t\varepsilon_{lm}\\
       &=\underbrace{\mathbf f_M^{(E)}\cdot\mathbf V_H}_{\text{\makebox[0pt]{work of its own force}}}\\
       &\quad+\underbrace{\tfrac12E_lE_m\left(\frac{\partial\varepsilon_{lm}}{\partial u_{ik}}\right)_{T,\mathbf E}
          \partial_tu_{ik}}_{\text{\makebox[0pt]{work of the third line of Eq.~\eqref{eq:helmholtz-stress-origins}}}}.
\end{aligned}
\label{eq:minkowski-abraham-power-deforming}
\end{equation}
The permittivity at a point of space changes for two reasons: matter of another permittivity is carried there with the velocity $\mathbf V_H$ of the body, and the matter there is strained. The first gives the work of the force of Eq.~\eqref{eq:minkowski-linear-electric-force}: here force and power belong together. The second gives the work of the third line, the electrostrictive stress that the tensor does not have~\cite[Sec.~4]{Robinson1975}. In an anisotropic solid the rotation adds the work of the tensor's own torque. \emph{For the linear dielectric the Minkowski tensor thus gives one load where the materials have many, and its energy balance contains the work of a force that it does not exert.}

\emph{The Abraham tensor.} Its energy density, its energy flux, and its power are those of the Minkowski tensor, also for the deforming body. Its stress is the symmetric part of Minkowski's, the first two lines of Eq.~\eqref{eq:helmholtz-stress-origins}, and its momentum density is $\mathbf g_{\rm vac}$, without Minkowski's added $\mathbf P\times\mathbf B$ (Table~\ref{tab:tensor-comparison}). For isotropic response in a static field its stress and its force are those of Eqs.~\eqref{eq:minkowski-static-stress} and \eqref{eq:minkowski-linear-electric-force}. The argument is therefore the same, step by step: all the energy is in the field, the stress of the tensor is the whole load, the third line is missing, and the energy balance contains its work.

One may add the third line to the two tensors, as for a fluid whose permittivity depends on its density~\cite[Secs.~15 and 75]{LandauLifshitzVol8}. The force is then the load on that fluid~\cite{MakarovRukhadze2009,LaiSuenYoung1981}, and the tensor contains its law: it has to be written anew for each material.

\subsubsection{Nonlinear and lossy response}
\label{subsubsec:audit-nonlinear}

Real dielectrics dissipate, and their response is nonlinear at high fields. The failure of the two tensors is then plainer than in the linear reversible case: it needs no deformation, and it shows in the energy balance of a body at rest.

\emph{The vacuum-form tensor.} Nothing changes with the response. Its force acts on the ions and on the bound electrons, its work is done on the electrons at their velocity $\mathbf W$, and binding and drag take it up (Sec.~\ref{subsec:lossy-rigid-dielectric-fecc}):
\begin{equation}
\begin{aligned}
    P_{\rm vac}
    &=\underbrace{\mathbf f_{{\rm L},{\rm ret},e}\cdot\mathbf W}_{\text{\makebox[0pt]{work of its own force}}}\\
    &=\underbrace{\partial_tu_{\rm bind}}_{\text{\makebox[0pt]{stored}}}
     \;+\;\underbrace{q_{\rm th}}_{\text{\makebox[0pt]{heat}}}.
\end{aligned}
\label{eq:vacuum-power-origins}
\end{equation}
The heat, $q_{\rm th}=-\mathbf f_{{\rm drag},e}\cdot\mathbf W\ge0$, is the work against the drag on the moving electrons, at every instant. The law of the material decides only how the work divides. \emph{For every response the power of the vacuum-form tensor is thus the work of its own force on the moving electrons, and the heat is a part of that work at every instant.}

\emph{The Minkowski tensor.} Its energy density stays $u_{\rm vac}+\tfrac12\mathbf E\cdot\mathbf P$ for every response. Beyond the energy of the retained fields, the energy flux delivers $\mathbf E\cdot\partial_t\mathbf P$ per unit volume and time, and the tensor holds $\partial_t(\tfrac12\mathbf E\cdot\mathbf P)$ of it as energy of the field. The rest is its power:
\begin{equation}
\begin{aligned}
    P_M
    &=\underbrace{\mathbf E\cdot\partial_t\mathbf P}_{\text{\makebox[0pt]{delivered}}}
      \;-\;\underbrace{\partial_t\!\left(\tfrac12\mathbf E\cdot\mathbf P\right)}_{\text{\makebox[0pt]{held in the field}}}\\
    &=\tfrac12\left(\mathbf E\cdot\partial_t\mathbf P-\mathbf P\cdot\partial_t\mathbf E\right).
\end{aligned}
\label{eq:minkowski-abraham-power}
\end{equation}
In familiar terms, $\mathbf E\cdot d\mathbf D$ is delivered and $d(\tfrac12\mathbf E\cdot\mathbf D)$ is held. The power vanishes while $\mathbf P$ follows $\mathbf E$ in fixed proportion. In Eq.~\eqref{eq:minkowski-abraham-power-deforming} the proportion changed because the body moved and deformed; here it changes in a body at rest, because the response is nonlinear or lags behind the field.

In mechanics the power is the force times the velocity of the mass on which it acts (Sec.~\ref{sec:microscopic-baseline}). The force of the tensor is taken to act on the body as a whole~\cite[Chap.~4]{PenfieldHaus1967}, and the body is at rest:
\begin{equation}
    P_M\;\neq\;\underbrace{\mathbf f_M\cdot\mathbf V_H}_{\text{\makebox[0pt]{work of its own force}}}=0 .
\label{eq:minkowski-power-no-work}
\end{equation}
The tensor thus has a power and no work. The electrons do move, and one can always invent forces on them that reproduce a given power: a pair of opposite forces on the two species changes the power without changing the force. This identifies nothing. The forces of the retained fields on the two species are not open to invention: they are given by Eq.~\eqref{eq:retained-species-lorentz-force}, and their work is $\mathbf E\cdot\partial_t\mathbf P$, not $P_M$.

Without losses, take the reversible law $P=\alpha E^3$ with constant $\alpha>0$. As the field rises from zero, $\tfrac34\alpha E^4$ is delivered and the tensor holds $\tfrac12\alpha E^4$. The remaining third has passed to the matter, by the balance of the tensor, and it returns in full when the field is removed. Matter takes energy from a field, and gives it back, as the work of a force on a moving mass. The tensor has no such mass: its force acts on a body at rest. In its account, energy leaves the field and returns to it, and nothing has moved.

With losses, part of what is delivered does not return. Over a cycle that returns the fields and the recoverable state of the material to their initial values, the energy of the tensor returns to its value, and its power sums to
\begin{equation}
    \oint P_M\,dt
    =\oint\mathbf E\cdot d\mathbf P>0,
\label{eq:minkowski-loss-cycle}
\end{equation}
the area of the hysteresis loop, which appears in the body as heat. The balance of the tensor thus contains the heat, and nothing that produces it: heat is the work of a drag on a moving mass, and the tensor has neither. Only the sum over the closed cycle is the heat; within the cycle the power also contains energy that returns, as in the case without losses. \emph{For nonlinear and lossy response the Minkowski tensor thus fails the simplest test of mechanics, even for a body at rest. Power is force times velocity; its force moves no mass, and its power is not zero.}

\emph{The Abraham tensor.} Its energy density and its power are those of the Minkowski tensor, and its force acts on the same body at rest. The result is the same.

One may let the two tensors hold all that is delivered, $\int\mathbf E\cdot d\mathbf D$ in place of $\tfrac12\mathbf E\cdot\mathbf D$, and the power then vanishes for reversible response. The energy entry is then $u_{\rm vac}+u_{\rm bind}$, a sum that has to be written anew for each material and so contains its law (Sec.~\ref{subsec:rigid-reversible-dielectric-fecc}). With losses the integral is no function of state: it grows by the area of the loop in every cycle, and the heat would be counted as energy of the field.

\subsection{The result: the division of the total is not arbitrary}
\label{subsec:audit-protocol}
\label{subsubsec:audit-result}

\emph{The contrast.} Table~\ref{tab:tensor-comparison} sets the three tensors side by side. The vacuum-form tensor contains the polarization only as the charge and the current that it represents. Its force is the Lorentz force of the retained fields on the ions and the bound electrons, and its power is the work of that force: it answers the four questions for every response, and no law of the material enters. With that law added, its two balances lead to the heat at every instant and to the whole load on the body, for the linear dielectric the complete Helmholtz stress (Sec.~\ref{subsec:mechanical-load-closure}). The Minkowski and Abraham tensors take the polarization into the tensor as a field of its own. This builds a law of the material into them, the energy of a linear dielectric and one rule of transmission, and their two balances are no longer a force on the bound charge and the work of that force. They give no answer to the four questions that holds for every response, and even for the linear dielectric their stress lacks the line of the Helmholtz stress that depends on the material. Neither tensor divides the total between the field and the matter.

\emph{The objection.} For the most widely held resolution of the debate, such a verdict has no consequence: it regards the division of the total energy--momentum tensor into an electromagnetic and a material part as arbitrary, a convention. Every electromagnetic tensor has a material counterpart with which it sums to the same total, the predictions of all such pairs agree, and the choice among them is a matter of preference~\cite[Secs.~VI and XI]{Pfeifer2007}, \cite[Secs.~7.6 and 7.7]{PenfieldHaus1967}, \cite[Chap.~V, Sec.~7]{DeGrootSuttorp1972}; see also Refs.~\cite{Kranys1979EquivalenceAbrahamMinkowski,Kranys1982NonUniqueness,Maugin1980EquivalenceAbrahamMinkowskiOthers,Barnett2010,MilonniBoyd2010,Silveirinha2017ReexaminationAbrahamMinkowski}. In its sharpest form the claim is that only the total tensor has a clear physical meaning~\cite{RamosRubilarObukhov2015FirstPrinciples}, or that the total momentum does not separate into field and matter parts~\cite{Crenshaw2013}.

\emph{What the claim establishes.} A term moved from the field to the matter leaves the sum unchanged:
\begin{equation}
\begin{aligned}
    \mathcal T_{\rm tot}
    &=\mathcal T_{\rm vac}+\mathcal T_{\rm mat}\\
    &=(\mathcal T_{\rm vac}+\Delta\mathcal T)
      +(\mathcal T_{\rm mat}-\Delta\mathcal T),
\end{aligned}
\label{eq:arbitrary-split-ledger}
\end{equation}
where $\mathcal T_{\rm mat}$ stands for the rest of the total, Eq.~\eqref{eq:complete-account}. We do not dispute this: it is an identity of sums. It holds for every $\Delta\mathcal T$, even for $\Delta\mathcal T=-\mathcal T_{\rm vac}$, which leaves the field without energy or momentum, and no observation can contradict it~\cite[Sec.~VI.A]{Pfeifer2007}.

\emph{The same step for a free charge.} For a free charge in vacuum, with $m\dot{\mathbf v}=q(\mathbf e+\mathbf v\times\mathbf b)$, the same step adds a term to both sides:
\begin{equation}
    m\dot{\mathbf v}+\Delta\mathbf f
    =q(\mathbf e+\mathbf v\times\mathbf b)+\Delta\mathbf f .
\label{eq:split-free-charge}
\end{equation}
The equation holds for every $\Delta\mathbf f$ and predicts the same motion. But its right side is no longer the force of the field on the charge, and its left side is no longer the rate of change of the momentum of the charge; it cannot even be written from the charge alone. No one calls it a matter of choice what force a field exerts on a free charge. An equation of motion is more than an equality: each of its terms stands for something, and the identity of sums gives this up.

\emph{The same step in matter.} The balances of the two continua, Eq.~\eqref{eq:retained-carrier-host-pair}, have the same form: on the left the mass of the ions and the electrons and its motion, with the host forces on it; on the right the force of the retained fields on their charge, and its work. A candidate adds $\Delta\mathbf f_X$ and $\Delta P_X$, Eq.~\eqref{eq:residual-audit}, on the right, and its material counterpart adds them on the left:
\begin{equation}
\begin{aligned}
    \boldsymbol{\mathcal K}_{ie}^{\rm ret}-\mathbf f_{\rm host}+\Delta\mathbf f_X
       &=\mathbf f_{\rm vac}+\Delta\mathbf f_X,\\
    \mathcal E_{ie}^{\rm ret}-P_{\rm host}+\Delta P_X&=P_{\rm vac}+\Delta P_X .
\end{aligned}
\label{eq:split-two-continua}
\end{equation}
Again the equality holds, and again neither side is what it was. The right sides contain what the candidate adds, which belongs to the matter or to no part of the total. What belongs to the matter follows a law of the material, and a fixed expression in the fields can contain only one such law: for every other material the candidate fails the test, as the Minkowski and Abraham tensors do (Sec.~\ref{subsec:audit-minkowski}). The force density of Einstein and Laub~\cite{EinsteinLaub1908}, with any tensor that reproduces it~\cite[Sec.~3.4]{AnghinoniEtAl2022}, is of this kind: it is an expression in the fields and $\mathbf P$ alone, and bodies with the same retained fields and polarization carry different loads~\cite[Sec.~V\,C]{SchulteWesthoffLocalLoads2026}, \cite[Sec.~6.4]{SchulteWesthoffDielectricEnergy2026}, \cite{SunWangNgZhouChan2015,WangNgXiaoChan2016} (Sec.~\ref{subsec:information-loss-tensor}). What belongs to no part gives a force on no mass (Sec.~\ref{subsubsec:canonical-local-tensor-test}). This holds whatever the candidate adds. The left sides are the balance of no mass: the material part $\mathcal T_{\rm mat}-\Delta\mathcal T$ is defined only as what remains of the total~\cite[Sec.~VI]{Pfeifer2007}, on whose form, by the same review, no agreement has been reached~\cite[Secs.~VI.A and XI]{Pfeifer2007}.

\emph{What is fixed, and what remains open.} The division is therefore not arbitrary. Once the averaging that defines $\mathbf E$ and $\mathbf B$ is chosen, it is fixed, up to terms that change neither balance; and in smooth bulk the separation of scales makes these fields insensitive to that choice~\cite[Sec.~4.2]{SchulteWesthoffDielectricEnergy2026}. What the matter does with the force and the work it receives is the subject of material laws: how the work divides~\cite{SchulteWesthoffDielectricEnergy2026}, and how the load reaches the body~\cite{SchulteWesthoffLocalLoads2026}. They are open to experiment, and they are not open to choice.

\subsection{Canonical and kinetic momentum interpretations}
\label{subsec:canonical-kinetic}

No experiment measures the momentum of light in a medium. Each measures a force or a torque on a body, or the displacement or deformation that results~\cite[Sec.~III]{Pfeifer2007}; a momentum is inferred from it with an assumption about the surrounding matter~\cite[Sec.~7]{MilonniBoyd2010}. At the resolution of the carriers every such outcome has one cause, the Lorentz force on ions and electrons, and the analyses of the experiments start from it~\cite{Gordon1973,ObukhovHehl2003EnergyMomentumForcesMatter,BarnettLoudon2006ForceDielectricMedium,HindsBarnett2009,Mansuripur2010,MilonniBoyd2010}. There is one field at this resolution, $\mathbf e$ and $\mathbf b$, and this force is the momentum balance of its vacuum-form tensor (Sec.~\ref{sec:microscopic-baseline}). Nothing is left to choose.

The second resolution identifies the Abraham and the Minkowski momentum of light with its kinetic and its canonical momentum, so that both are correct, each in its own role~\cite{Barnett2010,BarnettLoudon2010EnigmaOpticalMomentum}. The identification rests on a single atom: the kinetic and the canonical momentum of an atom in a light field differ by the same amount as the two integrated momenta of the light~\cite{HindsBarnett2009}. We ask whether this correspondence identifies either tensor as that of the electromagnetic field.

Two things limit what the argument can settle, whatever its outcome. The first is the scale. The Abraham and Minkowski tensors are statements about the macroscopic fields in matter; the argument is made for one atom and carried to a medium by adding the momenta of its dipoles~\cite{Barnett2010}. For one atom, as the authors note themselves, both momenta are obtained by applying the same Lorentz force~\cite[Sec.~3]{BarnettLoudon2010EnigmaOpticalMomentum}: the argument has started from the vacuum-form tensor. The second is the object. A momentum is one entry of four, and the force of a tensor needs its stress as well, Eq.~\eqref{eq:mic-fecc-general}: a tensor has to be judged as a whole.

We therefore move to the scale of the argument: a resolved pair of an ion and an electron with its unaveraged fields, whose charges a microscopic polarization field $\mathbf p$ represents exactly, without averaging and without added matter~\cite[Sec.~4.3]{SchulteWesthoffDielectricEnergy2026}. We first take the two tensors as wholes, then follow the momentum argument itself, from a single carrier to the dipole. Section~\ref{subsec:information-loss-tensor} returns to the macroscopic resolution and to the individual experiments.

\subsubsection{Local force test with resolved charges}
\label{subsubsec:canonical-local-tensor-test}

The kinetic--canonical argument applies the two momentum densities to a single atom, with a polarization concentrated at the position of the atom~\cite[Eq.~(1)]{HindsBarnett2009}. Its correspondence concerns integrated momenta and leaves open where the force acts inside the atom; a tensor of the field has to say this as well. We therefore keep the ion and the electron resolved, with their masses and the Lorentz force on each. The example shows that an equal total force can go together with a force where no mass is present.

Consider an ion of charge $q$ and an electron of charge $-q$, held at a fixed separation, with smooth localized profiles and an empty gap between them. Their charge density has the exact representation $\rho_{\rm mic}=-\nabla\cdot\mathbf p$. In the construction of \ref{app:separated-charge-support}, $\mathbf p$ extends through the gap, although its divergence and the mass densities of both carriers vanish there.

We evaluate all tensor expressions on the same unaveraged fields, which include the field of the pair and the applied field, with $\mathbf p$ in place of $\mathbf P$. The configuration is static, so the rates of change of the momentum densities vanish and the force of each tensor is the divergence of its stress. Using Eq.~\eqref{eq:minkowski-linear-electric-force}, the vacuum-form and the Minkowski expressions give
\begin{equation}
\begin{aligned}
    \mathbf f_L^{(E)}
      &=\rho_{\rm mic}\mathbf e
       =-(\nabla\cdot\mathbf p)\mathbf e,\\
    \mathbf f_M^{(E)}
      &=(\mathbf p\cdot\nabla)\mathbf e
        -\tfrac12\nabla(\mathbf e\cdot\mathbf p).
\end{aligned}
\label{eq:mic-dipole-force-comparison}
\end{equation}
The Lorentz force vanishes in the empty gap. The Minkowski expression also depends on variations of the field and of the polarization, which need not vanish there. \ref{app:separated-charge-support} gives an explicit static construction in which both the Minkowski and the Abraham expressions produce force in parts of the gap.

The mismatch is immediate: no ion and no electron is there to receive this force. Read as forces of the field on matter, the two expressions require a mass that the system does not contain. They do not describe the electromagnetic load on this resolved pair. The force that they add also depends on how the charges are represented. Adding a static solenoidal field to $\mathbf p$ changes neither $\rho_{\rm mic}$ nor $\mathbf j_{\rm mic}$, but it changes the force of the two expressions, which contain $\mathbf p$ itself. The Lorentz force contains only the charge and the current.

Both expressions nevertheless give the same total force on the pair as the Lorentz force, because their stresses agree with the vacuum-form stress on a surface that encloses the pair outside the support of $\mathbf p$. A compensating material stress can restore the local balance, as discussed in Sec.~\ref{subsec:audit-protocol}. Neither the compensation nor the equal total makes the added force a force of the field on the charges of the pair.

\emph{The four questions at this scale.} The four questions apply here as they do in the continuum, and the outcome is starker. For the vacuum-form tensor the answers are those of Sec.~\ref{subsec:fecc}: the force acts on the ion and on the electron, each where its charge is, and its work is done at the velocity of that carrier. For the Minkowski and Abraham tensors: (i) The added force has no mass to act on. In the continuum of Sec.~\ref{subsec:audit-minkowski} it could still be read as a force on the body as a whole; in the gap there is no body. (ii) Its work has no velocity. If the pair is held fixed and the applied field varies slowly, no carrier moves and the Lorentz power vanishes everywhere. Both tensors nevertheless assign the power $-\tfrac12\mathbf p\cdot\partial_t\mathbf e_{\rm ext}$ in the gap, Eq.~\eqref{eq:mic-dipole-gap-power}: a power where nothing moves. (iii) Nothing in the gap balances the force, and (iv) nothing there takes up the power. The momentum correspondence that follows has to be read with this in view.

\subsubsection{Carrier dynamics and canonical momentum}
\label{subsubsec:canonical-carrier-dynamics}

The momentum of carriers and field is the reference for the canonical variables. In the nonrelativistic description of Sec.~\ref{sec:microscopic-baseline}, its density and its integral over all space are
\begin{equation}
\begin{aligned}
    \mathbf g_{\rm tot}^{\rm mic}
      &=\sum_a\rho_{m,a}\mathbf v_a
        +\varepsilon_0\mathbf e\times\mathbf b,\\
    \mathbf P_{\rm tot}
      &=\int_{\mathbb R^3}\mathbf g_{\rm tot}^{\rm mic}\,dV.
\end{aligned}
\label{eq:mic-total-momentum-canonical-reference}
\end{equation}
Equation~\eqref{eq:mic-total-momentum-canonical-reference} is the whole momentum of the system: the kinetic momentum of every carrier and the momentum of the microscopic fields $\mathbf e$ and $\mathbf b$, evaluated with the same regularization. The canonical variables introduced below change how this total is written, not what it contains.

For one carrier in prescribed smooth external potentials, with no other force, the minimally coupled Lagrangian is~\cite{Jackson1999}
\begin{equation}
    L_a=\tfrac12m_a\mathbf v_a^2
          +q_a\mathbf v_a\cdot\mathbf A_{\rm ext}-q_a\phi_{\rm ext},
\label{eq:mic-particle-canonical-kinetic}
\end{equation}
with the potentials taken at the position of the carrier. Its Euler--Lagrange equation is the Lorentz equation of motion,
\begin{equation}
    m_a\dot{\mathbf v}_a
      =q_a\left(\mathbf e_{\rm ext}
                 +\mathbf v_a\times\mathbf b_{\rm ext}\right),
    \label{eq:mic-particle-lorentz-dynamics}
\end{equation}
and its canonical and kinetic momenta are
\begin{equation}
\begin{aligned}
    \mathbf p_{a,\rm can}
      &=\frac{\partial L_a}{\partial\mathbf v_a}
       =m_a\mathbf v_a+q_a\mathbf A_{\rm ext},\\
    \mathbf p_{a,\rm kin}&=m_a\mathbf v_a.
\end{aligned}
    \label{eq:mic-particle-momentum-definitions}
\end{equation}

The term $q_a\mathbf A_{\rm ext}$ depends on the gauge. It carries the coupling to the field inside the canonical variable. The Lagrangian describes the same Lorentz dynamics, and the canonical momentum adds nothing to the momentum of Eq.~\eqref{eq:mic-total-momentum-canonical-reference}. The tensor of the field remains the vacuum-form tensor, with the momentum density $\varepsilon_0\mathbf e\times\mathbf b$. We next follow the same distinction for the common motion of the two carriers of the pair.

\subsubsection{Point-dipole dynamics and the momentum correspondence}
\label{subsubsec:canonical-dipole-correspondence}

The derivation keeps three things apart: what is exact for the extended pair, where the point-dipole approximation enters, and what it drops. Consider again the resolved, neutral ion--electron pair of Sec.~\ref{subsubsec:canonical-local-tensor-test}. A localized microscopic polarization field $\mathbf p$ represents both sources, with no separate magnetization current:
\begin{equation}
\begin{aligned}
    \rho_{\rm mic}&=\rho_i+\rho_e=-\nabla\cdot\mathbf p,\\
    \mathbf j_{\rm mic}&=\rho_i\mathbf v_i+\rho_e\mathbf v_e
      =\partial_t\mathbf p.
\end{aligned}
    \label{eq:dipole-force-source-representation}
\end{equation}
Here $\rho_i,\rho_e$ and $\mathbf v_i,\mathbf v_e$ are the charge densities and the velocities of the two carriers. \ref{app:separated-charge-support} gives an explicit construction of this representation.

For a fixed volume $\mathcal U$ that encloses the charges, the currents, and the support of $\mathbf p$ throughout the interval considered, the electric dipole moment of the pair is
\begin{equation}
\begin{aligned}
    \mathbf d(t)
      &\equiv\int_{\mathcal U}\mathbf x\rho_{\rm mic}\,dV\\
      &=-\int_{\mathcal U}\mathbf x(\nabla\cdot\mathbf p)\,dV
       =\int_{\mathcal U}\mathbf p\,dV.
\end{aligned}
\label{eq:dipole-moment-source-integrals}
\end{equation}
The last equality follows by integration by parts, with $\mathbf p=\mathbf0$ on $\partial\mathcal U$. This moment is defined for the extended pair, without a point-dipole approximation. Different distributions of charge can have the same $\mathbf d$, so the moment alone does not determine the distribution of charge inside the pair. Differentiating the integral of the polarization gives
\begin{equation}
    \dot{\mathbf d}
      =\int_{\mathcal U}\partial_t\mathbf p\,dV
      =\int_{\mathcal U}\mathbf j_{\rm mic}\,dV.
    \label{eq:dipole-moment-current-integral}
\end{equation}

To connect the dynamics of the resolved carriers with the Lagrangian of the dipole, we first calculate the total Lorentz force that smooth external fields exert on the pair. It drives the motion of its center of mass. Using Eq.~\eqref{eq:dipole-force-source-representation}, we obtain
\begin{equation}
\begin{aligned}
    \mathbf F_{\rm ext}
      &=\int_{\mathcal U}
        \left[-(\nabla\cdot\mathbf p)\mathbf e_{\rm ext}
              +\partial_t\mathbf p\times\mathbf b_{\rm ext}\right]dV\\
      &=\int_{\mathcal U}
        \left[(\mathbf p\cdot\nabla)\mathbf e_{\rm ext}
              +\partial_t\mathbf p\times\mathbf b_{\rm ext}\right]dV.
\end{aligned}
\label{eq:dipole-integrated-lorentz-force}
\end{equation}
The last equality follows by integration by parts; the surface term vanishes because $\mathbf p=\mathbf0$ on $\partial\mathcal U$. The integrand of the second line is the force density on dipoles used in the literature~\cite{BarnettLoudon2006ForceDielectricMedium,MansuripurZakharianWright2013,Markel2023ElectrostaticForceDensity}. It is the Lorentz force on the two charges in another form, and the two lines agree as integrals over the pair, not point by point.

Up to this point, the force integral and the identities for the dipole moment are exact. We now take the gradient of the external electric field, $\nabla\mathbf e_{\rm ext}$, and the magnetic field $\mathbf b_{\rm ext}$ as constant across the pair, with their values at its center of mass $\mathbf R$. These factors can then be taken outside the integrals:
\begin{equation}
\begin{aligned}
    \mathbf F_{\rm ext}
      &\simeq\left[\left(\int_{\mathcal U}\mathbf p\,dV\right)
          \cdot\nabla\right]\mathbf e_{\rm ext}(\mathbf R,t)\\
      &\quad+\left(\int_{\mathcal U}\partial_t\mathbf p\,dV\right)
          \times\mathbf b_{\rm ext}(\mathbf R,t).
\end{aligned}
    \label{eq:dipole-force-factorization}
\end{equation}
The two source integrals are $\mathbf d$ and $\dot{\mathbf d}$ of Eqs.~\eqref{eq:dipole-moment-source-integrals} and \eqref{eq:dipole-moment-current-integral}. This point-dipole approximation keeps the leading electric-dipole force; magnetic-dipole and higher multipole terms are omitted. Let both carriers move under conservative internal binding, with no external mechanical force, and write $M=m_i+m_e$, $\mathbf V=\dot{\mathbf R}$. At an instant with $\mathbf V=\mathbf0$, the equation of translational motion is therefore~\cite[Eq.~(2)]{HindsBarnett2009}
\begin{equation}
\begin{aligned}
    M\dot{\mathbf V}
      =\mathbf F_{\rm ext}
      &\simeq(\mathbf d\cdot\nabla)\mathbf e_{\rm ext}(\mathbf R,t)\\
      &\quad+\dot{\mathbf d}\times\mathbf b_{\rm ext}(\mathbf R,t).
\end{aligned}
    \label{eq:dipole-same-lorentz-dynamics}
\end{equation}
This is the Lorentz force from which the kinetic--canonical argument starts. It is the force on the pair as a whole and does not say where in the pair the force acts. The internal moment $\mathbf d(t)$ remains dynamical; its term $\dot{\mathbf d}\times\mathbf b_{\rm ext}$ is kept.

At this order, Faraday's law brings the same equation, at the same instant of rest, into the form
\begin{equation}
    \frac{d}{dt}\left(M\mathbf V-\mathbf d\times\mathbf b_{\rm ext}\right)
      =\nabla_{\mathbf R}\left(\mathbf d\cdot\mathbf e_{\rm ext}\right).
    \label{eq:dipole-canonical-force-form}
\end{equation}
Here the fields are taken at $\mathbf R$, and the spatial derivative holds $\mathbf d$ fixed without freezing its change in time. The corresponding electric-dipole Lagrangian at this order is~\cite[Eqs.~(3)--(5)]{SonnleitnerBarnett2017Rontgen}
\begin{equation}
\begin{aligned}
    L&=\tfrac12M\mathbf V^2+L_{\rm int}
       +\mathbf d\cdot\mathbf e_{\rm ext}\\
     &\quad-\mathbf V\cdot(\mathbf d\times\mathbf b_{\rm ext}),
\end{aligned}
\label{eq:dipole-effective-lagrangian}
\end{equation}
where $L_{\rm int}$ describes the internal motion and the conservative binding and does not depend explicitly on $\mathbf R$ or $\mathbf V$. The canonical momentum conjugate to $\mathbf R$ and the kinetic momentum are therefore
\begin{equation}
\begin{aligned}
    \mathbf p_{\rm can}
      &=\frac{\partial L}{\partial\mathbf V}
       =M\mathbf V-\mathbf d\times\mathbf b_{\rm ext},\\
    \mathbf p_{\rm kin}&=M\mathbf V.
\end{aligned}
\label{eq:dipole-canonical-kinetic}
\end{equation}
The Euler--Lagrange equation $d(\partial L/\partial\mathbf V)/dt=\nabla_{\mathbf R}L$, evaluated at $\mathbf V=\mathbf0$ after taking the derivatives, returns Eq.~\eqref{eq:dipole-canonical-force-form}. That equation is therefore the rate of change of the canonical momentum, while $M\dot{\mathbf V}$ is the force on the mass of the pair.

For the nonmagnetic tensor expressions, $\mathbf g_M^{\rm mic}-\mathbf g_A^{\rm mic}=\mathbf p\times\mathbf b$. In the same point-dipole approximation, the part of their integrated difference that is due to the external field equals the difference between the kinetic and the canonical momentum of the dipole~\cite{HindsBarnett2009,Barnett2010}:
\begin{equation}
\begin{aligned}
    \Delta\mathbf P_{MA}^{\rm ext}
      &\equiv\int_{\mathcal U}\mathbf p\times\mathbf b_{\rm ext}\,dV\\
      &\simeq\mathbf d\times\mathbf b_{\rm ext}(\mathbf R,t)\\
      &=\mathbf p_{\rm kin}-\mathbf p_{\rm can}.
\end{aligned}
\label{eq:dipole-integrated-momentum-correspondence}
\end{equation}
The full difference of the field momenta contains in addition the contribution of the self-field of the pair.

With Eq.~\eqref{eq:mic-total-momentum-canonical-reference}, the momentum of the pair and the field can thus be written in two ways~\cite[Eq.~(14)]{HindsBarnett2009}:
\begin{equation}
\begin{aligned}
    \mathbf P_{\rm tot}
      &=\mathbf p_{\rm kin}
        +\int\varepsilon_0\mathbf e\times\mathbf b\,dV\\
      &\simeq\mathbf p_{\rm can}
        +\int\left(\varepsilon_0\mathbf e\times\mathbf b
          +\mathbf p\times\mathbf b_{\rm ext}\right)dV .
\end{aligned}
\label{eq:dipole-two-splits}
\end{equation}
The first line is the momentum of the two carriers and the momentum of the field. The second moves the term $\mathbf d\times\mathbf b_{\rm ext}$ from the atom to the light. Hinds and Barnett describe it in the same way: dropping this term from the force on the atom adds it to the momentum of the light and turns the Abraham value into the Minkowski value~\cite{HindsBarnett2009}. The review of Pfeifer \emph{et al.} describes the same for a transparent rod traversed by a wave packet: the Minkowski approach counts the momentum of the moving rod as electromagnetic momentum~\cite[Sec.~VII]{Pfeifer2007}. This is the shift of Sec.~\ref{subsec:audit-protocol}, made for one term of one entry. The momentum that the Lorentz force leaves with the light is $\varepsilon_0\mathbf e\times\mathbf b$, the entry of the vacuum-form tensor, with which the Abraham expression for the momentum density coincides in nonmagnetic matter (Table~\ref{tab:tensor-comparison}).

\emph{What is established, and what does not follow.} Equation~\eqref{eq:dipole-integrated-momentum-correspondence} is established. It relates differences of integrated momenta. The relation rests on the point dipole, a description at lower resolution like the averaged fields of Sec.~\ref{subsec:field-split}: the total force on the pair is kept, and where in the pair it acts is given up. It does not identify either the Minkowski or the Abraham tensor as the tensor of the field of the resolved pair. Section~\ref{subsubsec:canonical-local-tensor-test} has shown the mismatch: taken as wholes, both tensors put force in the empty gap, where no mass is present to receive it. The correspondence of the integrated momenta does not remove this force.

Canonical variables change neither the vacuum-form tensor nor the force and the work that its two balances are. Nor does the correspondence say anything about the macroscopic fields. Their tensor, with its four entries and its two balances, was identified in Sec.~\ref{sec:macroscopic-compression}, and the correspondence supplies no other.

\emph{Canonical and metric definitions.} A tensor can also be defined from a Lagrangian, by Noether's theorem or by variation of the metric. For the Lagrangian of the macroscopic field in a linear medium, the first gives the Minkowski tensor~\cite[Sec.~5.1]{Obukhov2008} and the second, for a medium that is homogeneous and isotropic at rest, the Abraham tensor~\cite{AntociMihich1997}. This Lagrangian contains the permittivity, and with it the law of the material~\cite{Philbin2011DispersiveEnergyMomentum}. Once the bound charge is treated as charge, the canonical balance is that of the vacuum-form tensor, with the Lorentz force on all charge~\cite[Eqs.~(31)--(34)]{Obukhov2008}.

\subsection{What the experiments measure}
\label{subsec:information-loss-tensor}

What the experiments of the debate measured is a force (Sec.~\ref{subsec:canonical-kinetic}), at the resolution of the carriers the Lorentz force, from which both momenta of the kinetic--canonical resolution follow: an experiment that agrees with either momentum agrees with this force. At the macroscopic resolution this force is the force of the vacuum-form tensor of $\mathbf E$ and $\mathbf B$, and it reaches what is measured in two ways (Sec.~\ref{subsec:mechanical-load-closure}). The total force and torque on a body follow from $\mathbf E$ and $\mathbf B$ alone, for every material; a deformation follows together with the binding energy of the material, hence with its law. The measurements of the debate are of these two kinds.

\emph{Totals.} For a body in vacuum the total forces of the tensors differ only by the term $d/dt\int\mathbf P\times\mathbf B\,dV$, which the Minkowski force lacks (Table~\ref{tab:tensor-comparison}). The torque that Walker and co-workers measured on a dielectric disk in slowly varying fields~\cite{WalkerLahozWalker1975,WalkerWalker1977} is that of the Lorentz force on the bound charge and its current, $-(\nabla\cdot\mathbf P)\mathbf E+\partial_t\mathbf P\times\mathbf B$~\cite[Sec.~4]{ObukhovHehl2003EnergyMomentumForcesMatter}; the second term is the force that Einstein and Laub found missing from Minkowski's expression~\cite{EinsteinLaub1908}. The torque had been read as support for the Abraham tensor~\cite[Sec.~IV.C]{Pfeifer2007}. For a body immersed in a medium in steady light, the pressure on a mirror in a liquid~\cite{JonesRichards1954,JonesLeslie1978} and the recoil of an atom in a dilute gas~\cite{CampbellEtAl2005} are proportional to the index $n$~\cite[Sec.~4]{MilonniBoyd2010}. In a traveling wave in the medium the magnetic field is $n/c$ times the electric field, so that the Lorentz force on a transverse current is $\mathbf J\times\mathbf B=(n/c)(\mathbf J\cdot\mathbf E)\hat{\mathbf k}$, with $\hat{\mathbf k}$ the direction of propagation: the force is $n/c$ times the power, for the same carriers, as in the recoil of an absorbing atom~\cite[Eq.~(76)]{MilonniBoyd2010} and in the photon drag on the free carriers of a semiconductor~\cite{GibsonEtAl1980,LoudonBarnettBaxter2005}. On a mirror the pressure is proportional to $n$ for a perfect electric conductor and to $1/n$ for a perfect magnetic one~\cite[Sec.~3.3.3]{Kemp2015MacroscopicTheoryOpticalMomentum}. The factor belongs to the field at the surface of the body, not to a momentum that arrives with the light.

\emph{Deformation.}\label{subsubsec:dipole-model-force-prediction} The bulge of a liquid surface under a laser beam~\cite{AshkinDziedzic1973} has been explained through radial forces on the liquid~\cite{Gordon1973}. It tests the local load on the liquid~\cite{Brevik1979}; whether the surface rises or sinks depends on the flow that the light drives in the liquid~\cite{Leonhardt2014,ZhangShePengLeonhardt2015,VermaBrevikLi2025}. This load follows from the force of the retained fields and from the binding energy of the liquid under deformation; for a linear liquid it is the Helmholtz force~\cite[Sec.~15]{LandauLifshitzVol8}, the form of Eq.~\eqref{eq:helmholtz-stress-three-parts} for a permittivity that depends on density. A force density written for a specified material states this outcome for that material~\cite{LaiSuenYoung1981,Markel2023ElectrostaticForceDensity}. The force on charges bound in dipoles, $(\mathbf P\cdot\nabla)\mathbf E_{\rm ext}$ for an external field $\mathbf E_{\rm ext}$ that is smooth across them, is the electric-dipole term of the Einstein--Laub force~\cite{EinsteinLaub1908}, \cite[Sec.~V\,A]{SchulteWesthoffLocalLoads2026}, which holds in integrated form for classical dipolar media under stated assumptions~\cite[Secs.~3.2 and~4]{Yaghjian2023ForceMomentumDipolarMedia}. For a medium whose susceptibility is proportional to its density, the Helmholtz force is this dipole force; where the two differ, as in a dense liquid, the measurements of electrostriction follow the Helmholtz force~\cite[Secs.~2.3 and 2.4]{Brevik1979}. Such laws are statements about the matter, and they establish no other tensor of the field.

The experiments agree with the vacuum-form tensor, whose two balances are this force and its work. None of them singles out a tensor; the two requirements do.

\section{Conclusion}
\label{sec:conclusion}

An energy--momentum tensor of a field asserts a force on matter and the work of that force. We have held the tensors of the debate to two requirements of elementary mechanics, consistency and generality: the two balances must be a force on an identified mass and the work of that same force at the velocity of that mass, and this must hold for every material and every motion. Of the tensors written in the macroscopic fields, one remains: the vacuum-form tensor of $\mathbf E$ and $\mathbf B$. These fields are a part of the microscopic field, and the two balances of their tensor are their own share of the Lorentz force on the ions and the electrons and the work of that force at the velocities of these carriers. No law of the material enters, so this holds for linear and nonlinear, reversible and lossy response, at rest and in motion, and the tensor keeps its symmetric, covariant form in every inertial frame. For the electric response of matter in classical electrodynamics, this is the tensor of the macroscopic electromagnetic field.

The role of a tensor of the field is thereby defined. Its two balances are a budget: the force that the fields exert on the ions and the electrons, and the work of that force. What is measured follows from this budget. The total force and torque on a body follow from the fields alone. The load that deforms a body and the heat follow together with laws of the material~\cite{SchulteWesthoffDielectricEnergy2026,SchulteWesthoffLocalLoads2026}: the power records the energy passed to the matter, the part of it that the drag on the moving charges takes up is the heat, and the change of its stored part under deformation completes the load, which for a linear dielectric is the Helmholtz stress, electrostriction included. No force density written in the fields alone is this load, and a comparison of force densities could therefore not decide the debate. The laws of the material are open to experiment, and they are not open to choice.

With this budget, macroscopic electrodynamics, continuum mechanics, and thermodynamics, three descriptions of the same ions, electrons, and fields, form one map. Every part of the total energy and momentum has an owner, the macroscopic fields exchange with the rest through this budget alone, and every exchange is a force on an identified mass and its work. The structure is the same for every material, and the laws of the material supply the amounts: a coupled model takes the budget from the fields and adds these laws.

The Minkowski and Abraham tensors do not meet the requirements. They take the polarization as a field of its own. The fields and their sources do not fix it; a statement about the matter does, and what the two tensors add to the vacuum form is thereby a law of the material. Taken as wholes, their balances are not a force on charged mass and its work, for any response of the dielectric. For linear response they give one load where the materials have many, without electrostriction, and when the body deforms their power contains the work of a force that they do not exert. For nonlinear or lossy response they have a power while their force does no work, even in a body at rest. At the scale of resolved charges they put force and power into empty space. What any other candidate adds belongs likewise to the matter, or to no part of the total.

The two prevailing resolutions stay true in what they establish, and neither selects a tensor. The equivalence of the totals is an identity of sums and says nothing about the parts: applied to a free charge, the same step would make the force of the field on that charge a matter of choice. The kinetic--canonical correspondence concerns one entry of four and starts from the Lorentz force. The experiments of the debate measured forces on bodies, which the Lorentz force explains.

The tensor and its identities are known, and no new equation or experiment is needed: the result follows from Maxwell--Lorentz theory, from which the analyses of the debate themselves start. What is new is the standard, applied to the charges themselves, on which the field acts, and what it decides: which tensor is that of the field, what its role is, and how it completes the map. The two requirements judge a tensor as a whole, by its two balances together; they are what force and work mean, not one more interpretation. The argument is classical and concerns the electric response of matter; magnetization, intrinsic spin, and fluctuation-induced forces are left to separate work.

Together with Refs.~\cite{SchulteWesthoffLocalLoads2026,SchulteWesthoffDielectricEnergy2026}, the result places three questions on one basis: which load deforms a body, where the work of the field goes, and which tensor is that of the field. In all three, the fields fix the budget and the totals, and the material fixes the rest. Once $\mathbf E$ and $\mathbf B$ are defined, the line between the two is not a convention.

\section*{CRediT authorship contribution statement}
\textbf{Bela Schulte Westhoff:} Conceptualization, Methodology, Formal analysis, Investigation, Software, Visualization, Writing -- original draft, Writing -- review \& editing.

\section*{Declaration of competing interest}
The author declares no known competing financial interests or personal relationships that could have appeared to influence the work reported in this paper.

\section*{Funding}
This research did not receive any specific grant from funding agencies in the public, commercial, or not-for-profit sectors.

\section*{Data availability}
No data was used for the research described in the article. The results are derived in the text and in the appendix.

\section*{Declaration of generative AI and AI-assisted technologies in the manuscript preparation process}
During the preparation of this work the author used large language models from OpenAI (GPT) and Anthropic (Claude) for assistance in drafting and revising the text, in writing the Python code that produces the figures, and in checking calculations. After using these tools, the author reviewed and edited the content as needed, verified the scientific content by working through the derivations and comparing with the literature, and takes full responsibility for the content of the published article.

\appendix
\section{Separated charges and the empty-gap test}
\label{app:separated-charge-support}

This appendix gives the explicit construction behind the test with resolved charges in Sec.~\ref{subsubsec:canonical-local-tensor-test}. We specify the ion and electron distributions, represent their sources exactly through polarization, and show how the Minkowski and Abraham expressions put force in the empty gap while preserving the integrated Lorentz force. We first allow the separation to vary to establish both charge and current identities; the force comparison then holds the pair fixed. A last step lets the applied field vary in time and gives the power that the two tensors assign.

Choose nonnegative, smooth, compactly supported profiles $h(x,y)$ and $k(z)$, centered at the origin, with
\begin{equation}
    \int h\,dx\,dy=1,
    \qquad \int k\,dz=1,
    \qquad {\rm supp}\,k\subset[-a,a].
    \label{eq:app-pair-profile-normalization}
\end{equation}
These smooth profiles replace point carriers by localized spatial distributions and are fixed throughout the comparison. Put the ion center at the origin and the electron center at $\ell(t)\hat{\mathbf z}$. Their number and mass densities are
\begin{equation}
\begin{aligned}
    n_i&=h(x,y)k(z),
    &n_e&=h(x,y)k(z-\ell),\\
    \rho_{m,i}&=m_i n_i,
    &\rho_{m,e}&=m_e n_e.
\end{aligned}
\label{eq:app-pair-carrier-profiles}
\end{equation}
With charges $q$ and $-q$, and velocities $\mathbf v_i=\mathbf0$ and $\mathbf v_e=\dot\ell\hat{\mathbf z}$, the total sources are
\begin{equation}
\begin{aligned}
    \rho_{\rm mic}&=q h(x,y)[k(z)-k(z-\ell)],\\
    \mathbf j_{\rm mic}&=-q\dot\ell h(x,y)k(z-\ell)\hat{\mathbf z}.
\end{aligned}
\label{eq:app-pair-exact-sources}
\end{equation}
At $\ell=0$ the charge profiles cancel while their masses add. An exact polarization representation of their subsequent separation is
\begin{equation}
    \mathbf p(\mathbf x,t)
       =-q h(x,y)\hat{\mathbf z}
          \int_0^{\ell(t)}k(z-s)\,ds.
    \label{eq:app-pair-exact-polarization}
\end{equation}
Differentiation gives the exact source identities
\begin{equation}
    \rho_{\rm mic}=-\nabla\cdot\mathbf p,
    \qquad
    \mathbf j_{\rm mic}=\partial_t\mathbf p.
    \label{eq:separated-pair-source-alias}
\end{equation}
The volume integral of $\mathbf p$ is the dipole moment $\mathbf d=-q\ell\hat{\mathbf z}$. Thus no averaging or independent polarization carrier has been introduced.

We use $\Delta$ to denote the difference between a candidate tensor entry and the vacuum entry, evaluated on the same unaveraged fields and sources. With vanishing magnetization, the Minkowski differences are
\begin{equation}
\begin{aligned}
    \Delta u_M^{\rm mic}&=\tfrac12\mathbf e\cdot\mathbf p,
    &\Delta\mathbf S_M^{\rm mic}&=\mathbf0,\\
    \Delta\mathbf g_M^{\rm mic}&=\mathbf p\times\mathbf b,\\
    \Delta\mathbf T_M^{\rm mic}
       &=\mathbf e\otimes\mathbf p
          -\tfrac12(\mathbf e\cdot\mathbf p)\mathbf I.
\end{aligned}
\label{eq:mic-dipole-minkowski-shift}
\end{equation}
The conventional Abraham stress increment is the symmetric part of $\Delta\mathbf T_M^{\rm mic}$~\cite[Eqs.~(40) and (42)]{Pfeifer2007}. In a static electric configuration, the powers of the two tensors and the time derivatives of their momentum densities vanish. Applying Eq.~\eqref{eq:residual-audit} to the added stress and using $\rho_{\rm mic}=-\nabla\cdot\mathbf p$ gives the force comparison in Eq.~\eqref{eq:mic-dipole-force-comparison}.

We now hold the separation fixed at $\ell>2a$ and take the fields to be static. In the open gap $a<z<\ell-a$, both number densities vanish, whereas the integral in Eq.~\eqref{eq:app-pair-exact-polarization} equals one. Consequently
\begin{equation}
    \rho_{m,i}=\rho_{m,e}=\rho_{\rm mic}=0,
    \qquad \mathbf p=-q h(x,y)\hat{\mathbf z}.
    \label{eq:app-pair-empty-gap}
\end{equation}
Keep this configuration fixed and superpose the uniform static applied field $\mathbf e_{\rm ext}=E_0\hat{\mathbf z}$. The self-field remains included and unchanged; $\delta_{\rm ext}$ denotes the change in a quantity upon adding this applied field while holding the carrier configuration fixed. External sources lie outside the comparison region. Equation~\eqref{eq:mic-dipole-force-comparison} gives $\delta_{\rm ext}\mathbf f_L^{(E)}=\rho_{\rm mic}\mathbf e_{\rm ext}=\mathbf0$ in the gap. There the resulting change in the Minkowski--vacuum stress difference is
\begin{equation}
    \delta_{\rm ext}\Delta\mathbf T_M^{\rm mic}
       =qE_0h
          \left(\tfrac12\mathbf I
                      -\hat{\mathbf z}\otimes\hat{\mathbf z}\right),
    \label{eq:app-pair-gap-stress}
\end{equation}
which is symmetric because the applied field and polarization are collinear. The Abraham stress has the same applied increment. Taking the divergence gives
\begin{equation}
\begin{aligned}
    \delta_{\rm ext}\mathbf f_L^{(E)}&=\mathbf0,\\
    \delta_{\rm ext}\mathbf f_M^{(E)}
      =\delta_{\rm ext}\mathbf f_A^{(E)}
      &=\tfrac12qE_0\nabla_\perp h.
\end{aligned}
    \label{eq:mic-dipole-gap-witness}
\end{equation}
Here $\nabla_\perp$ denotes the transverse gradient. A nonconstant smooth compact transverse profile has regions with nonzero gradient, so both tensors give a force increment where the mass densities of both carriers are exactly zero. Because the discrepancy varies with the applied field, it cannot be removed for all applied amplitudes by a fixed self-field contribution. The two tensors thus have no answer to the first of the four questions here, although the total force is unchanged.

Indeed, $\mathbf p$ and hence $\Delta\mathbf T_M^{\rm mic}$ vanish on any enclosing surface outside the compact polarization support. In this static test,
\begin{equation}
    \int_{\mathcal U}(\mathbf f_M^{(E)}-\mathbf f_L^{(E)})\,dV
      =\oint_{\partial\mathcal U}
          \Delta\mathbf T_M^{\rm mic}\cdot\mathbf n\,dA=0.
    \label{eq:app-pair-zero-resultant}
\end{equation}
The Abraham stress increment also vanishes on this surface, so its integrated force difference is zero as well. These equal total forces coexist with different local force densities. Neither rewriting the sources in terms of polarization nor calculating the total dipole force supplies a mass in the empty gap.

The same construction tests the power. Keep the pair fixed and let the applied field vary, $E_0=E_0(t)$, slowly enough that the magnetic field that accompanies the change can be neglected. No carrier moves, so $\mathbf j_{\rm mic}=\mathbf0$ and the Lorentz power $\mathbf j_{\rm mic}\cdot\mathbf e$ vanishes everywhere. By Eqs.~\eqref{eq:residual-audit} and \eqref{eq:mic-dipole-minkowski-shift}, the Minkowski tensor adds the power $-\partial_t(\tfrac12\mathbf e\cdot\mathbf p)$, and the Abraham tensor adds the same. With $\mathbf p$ fixed and the self-field static, the power of both tensors in the gap is
\begin{equation}
    P_M=P_A
      =-\tfrac12\mathbf p\cdot\partial_t\mathbf e_{\rm ext}
      =\tfrac12q\dot E_0h .
    \label{eq:mic-dipole-gap-power}
\end{equation}
Both tensors thus assign a power where no mass is present and where nothing moves.

\bibliographystyle{elsarticle-num}
\bibliography{Paper_macro_EM_tensorReferences}

\end{document}